\documentclass[12pt,preprint]{aastex}

\begin{document}

\title{On the Rapid Collapse and Evolution of Molecular Clouds}

\author{Bruce G. Elmegreen}
\affil{IBM Research Division, T.J. Watson
Research Center, P.O. Box 218, Yorktown Heights, NY 10598,
bge@watson.ibm.com}

\begin{abstract}
Stars generally form faster than the ambipolar diffusion time,
suggesting that several processes short circuit the delay and
promote a rapid collapse. These processes are considered here,
including turbulence compression in the outer parts of giant
molecular cloud (GMC) cores and GMC envelopes, GMC core formation in
an initially supercritical state, and compression-induced triggering
in dispersing GMC envelopes. The classical issues related to star
formation timescales are addressed: high molecular fractions, low
efficiencies, long consumption times for CO and HCN, rapid GMC core
disruption and the lack of a stable core, long absolute but short
relative timescales with accelerated star formation, and the slow
motions of protostars. We consider stimuli to collapse from changes
in the density dependence of the ionization fraction, the cosmic ray
ionization rate, and various dust properties at densities above
$\sim10^5$ cm$^{-3}$.  We favor the standard model of subcritical
GMC envelops and suggest they would be long lived if not for
disruption by rapid star formation in GMC cores. The lifecycle of
GMCs is illustrated by a spiral arm section in the Hubble Heritage
image of M51, showing GMC formation, star formation, GMC disruption
with lingering triggered star formation, and envelope dispersal.
There is no delay between spiral arm dustlanes and star formation;
the classical notion results from heavy extinction in the dust lane
and triggered star formation during cloud dispersal. Differences in
the IMF for the different modes of star formation are considered.
\end{abstract}


\section{Introduction}

Gas contraction during star formation can overcome magnetic forces
in either of two ways, by diffusing through a supporting field or by
overwhelming it with a greater force from self-gravity. If the
equilibrium supporting field is termed critical, then the first of
these is subcritical contraction and the second is supercritical. In
the standard model, clouds begin as subcritical throughout and spend
a relatively long time (e.g., $\geq10t_{dyn}$ for dynamical time
$t_{dyn}$) contracting by ambipolar diffusion until their cores
become supercritical, and then they spend a relatively short time
($1-2t_{dyn}$) collapsing to stars in the core (Mestel 1965; Nakano
\& Tademaru 1972; Mouschovias 1976; Shu 1983; Tomisaka, Ikeuchi, \&
Nakamura 1990; Li \& Nakamura 2002). While there is a large body of
literature on this subcritical to supercritical transition (see
reviews in Shu, Adams, \& Lizano 1987; Mouschovias 1991; McKee et
al. 1993; Mouschovias, Tassis, \& Kunz 2006), there is growing
evidence that much of star formation actually begins closer to the
supercritical state, bypassing the long diffusion time of the
standard model (e.g., Nakano 1998; Hartmann et al. 2001).

This new view is based in part on the observation of infalling
motions at large ($\sim0.1$ pc) radii (Tafalla et al. 1998; Williams
et al. 1999; Williams \& Myers 1999; Wu, Zhu, et al. 2005; Walsh,
Bourke, \& Myers 2006; Williams, Lee, \& Myers 2006), which are
expected for supercritical collapse (Basu \& Ciolek 2004) and for
supercritical collapse following fast inflows (Fatuzzo, Adams \&
Myers 2004). It is also based on the relatively short time scales
for star formation (Lee \& Myers 1999; Jijina, Myers \& Adams 1999;
Ballesteros-Paredes, Hartmann, \& V\'azquez-Semadeni 1999; Elmegreen
2000; Lee, Myers \& Tafalla 2001; Myers 2005; Furuya, Kitamura \&
Shinnaga, 2006; Kirk, Ward-Thompson \& Andr\'e 2005, 2007;
J{\o}rgensen et al. 2007; see review in Ballesteros-Paredes \&
Hartmann 2007). Regarding these short time scales, it is important
to distinguish between the duration of star formation once it begins
in a cloud core, which is relatively short even in the standard
model (Basu 1997; Tassis \& Mouschovias 2004), and the total
lifetime of the core, including the time prior to star formation. We
consider here that even the total lifetime of giant molecular cloud
(GMC) cores is relatively short, unlike in the standard model. The
primary evidence for this is the short duration of star formation in
cluster-forming cores (Sect. \ref{sect:smallfast}) combined with the
low fraction of GMCs in a dormant, pre-star formation stage. There
are no known examples of GMCs with potential cluster-forming cores
hovering for $\sim10t_{dyn}$ at subcritical masses while the
magnetic field passively diffuses away. There are also few examples
of the oblate shapes that are expected from slow-diffusion models
(e.g. Ryden 1996).

Observations of magnetic field strengths show supercritical (i.e.,
weak) values directly (Troland et al. 1996; Roberts, Crutcher \&
Troland 1997; Bourke et al. 2001; Matthews et al. 2005) or values
beyond (i.e., weaker than) supercritical (Crutcher 1999; Glenn,
Walker, \& Young, 1999; Uchida, Fiebig, \& G\"usten 2001; Brogan \&
Troland 2001). Indirect observations like bent or hour-glass field
line shapes have been taken as evidence for supercritical fields too
(e.g., Greaves, Holland \& Murray 1995; Holland et al. 1996; Lai et
al. 2002; Cortes \& Crutcher 2006). Near-critical field values are
also found in cloud cores (Bertoldi \& McKee 1992; Curran et al.
2004; Cortes \& Crutcher 2006), and subcritical values in cloud
envelopes (Cortes, Crutcher, \& Watson 2005), but there are few or
no subcritical values in GMC cores unless extreme orientations are
assumed (Crutcher 2007).

Perhaps the biggest driver of our changing view is the recognition
that supersonic turbulence is pervasive in the ISM. If a cloud is
ever in quasi-equilibrium, then the presence of supersonic
turbulence in addition to magnetic fields automatically implies
rapid evolution toward supercritical cores and star formation. The
turbulence always decays quickly, in $\sim t_{dyn}$ (Stone,
Ostriker, \& Gammie 1998; MacLow et al. 1998), and the magnetic
field, having formerly shared the equilibrium support with
turbulence, is suddenly alone and insufficient to prevent collapse.
If star-forming clouds are never in equilibrium, then their
evolution is rapid from the start. The same is true if molecular
clouds are super-Alfv\'enic (Padoan et al. 1998, 1999), which means
their turbulent speeds exceed their Alfv\'en speeds. The decay of
this turbulence will rapidly convert these clouds to supercritical,
leading to prompt collapse without a long diffusion stage.  The
remaining question is whether turbulence is regenerated during this
collapse to sustain the cloud core life. We suggest in Sections
\ref{sect:huff} and \ref{sect:dest} that it is not, at least where
high mass stars form. Compressible turbulence inside a cloud core
should not delay star formation but speed it up by increasing both
the mass-to-flux ratio and the dynamical rate in the compressed
regions (Sect. \ref{sect:comp}). The energy input also disrupts the
core and moves the remaining parts of it and the envelope to the
side where it forms more stars in another dynamical time (Sect.
\ref{sect:dest}).

Evidently, the standard model of slow diffusion followed by rapid
collapse has to be supplemented by two new modes of star formation
in which gas becomes supercritical rapidly, in only one or two
dynamical times following cloud formation. One of these new modes
applies on a star-by-star basis, following turbulence-enhanced
diffusion in compressed sheets and filaments in the cloud envelope
(Elmegreen 1993; Fatuzzo \& Adams 2002, Zweibel 2002; Heitsch et al.
2004; Fatuzzo, Adams, \& Myers 2004; Li et al. 2004; Li \& Nakamura
2004; Nakamura \& Li 2005; Kudoh \& Basu 2007). The other applies to
a whole cloud core following a history of near-critical gas buildup
and a brief diffusion phase ($t_{dyn}$ to $2t_{dyn}$) that converts
the core to supercritical (Ciolek \& Basu 2001).

We propose here that massive cloud cores are born close to the
critical condition. We make a distinction between rapid ($\sim
t_{dyn}$) GMC core evolution and slow ($>t_{dyn}$) GMC envelope
evolution. GMC envelopes are exposed to background radiation so they
have high ionization fractions, and they apparently begin their
lives in a subcritical state (e.g., Ciolek \& Mouschovias 1995;
Cortes, Crutcher \& Watson 2005), while GMC cores are heavily
shielded with low ionization fractions and they probably begin their
lives in a critical state.   In the absence of core star formation,
GMC envelopes should last for several dynamical times, but because
the envelopes form cores quickly, and the cores form highly
disruptive stars quickly, the envelopes are doomed along with their
cores to have relatively short lives.  This does not mean the
envelopes are completely destroyed, however; their pieces are
scattered and triggered to produce secondary generations of stars
later. Some shredded pieces of GMC envelopes have the properties of
diffuse clouds (e.g., Pan et al. 2005; Sect. \ref{sect:dest}).

The outline of this paper is as follows. Section \ref{sect:sc}
considers the rapid onset of supercritical conditions in GMC cores.
Section \ref{sect:time} reviews the evidence for rapid star
formation on large (\ref{sect:bigfast}) and small
(\ref{sect:smallfast}) scales, the rapid evolution of HCN cores
(\ref{sect:hcn}), the rapid dispersal of cluster-forming cores
(\ref{sect:huff}), the acceleration of magnetic diffusion in GMC
envelopes (\ref{sect:comp}), the possible enhancement of magnetic
diffusion in GMC cores (\ref{sect:magdiff}), and the slow motions of
protostars (\ref{sect:motions}). Section \ref{sect:dest} illustrates
the morphology of cloud formation, evolution, and destruction using
the Hubble Heritage image of M51, and contrasts the points that GMC
core evolution is supercritical and fast but GMC envelope evolution
is subcritical and slow. Finally, the implications of rapid star
formation for the IMF are reviewed in Section \ref{sect:imf}, where
differences between supercritical cluster cores and
turbulence-compressed GMC envelopes are suggested. A summary of the
results is in Section \ref{sect:sum}.

\section{Cloud Formation and the Onset of Critical Magnetic
Support} \label{sect:sc}

Diffuse (i.e., non-self-gravitating) clouds form by localized
compressions involving stellar pressures or supersonic turbulence
generated on larger scales. They also form by shredding GMCs. The
ISM is an active but relatively dark environment, so as long as the
energy input is pervasive and fast while the temperature is low,
shocks form easily and make diffuse clouds on dynamical time scales.
Numerous simulations illustrate this process in detail (de Avillez
\& Breitschwerdt 2005; Piontek \& Ostriker 2005). Diffuse clouds do
not necessarily evolve into star-forming clouds. Some apparently do
(Sect. \ref{sect:dest}), but most should disperse quickly in the
turbulent flow pattern (Heitsch et al. 2006).

Self-gravitating clouds begin as diffuse clouds in the sense that
their formation starts with a transition from non-self-gravitating
to self-gravitating gas.  This transition seems to be initiated most
often on a galactic scale, where independent processes like spiral
density waves, or directly related processes like swing-amplified
gas instabilities and magneto-Jeans instabilities (Kim, Ostriker, \&
Stone 2002; Kim 2007) provide the environment for self-gravity to
take hold. More localized compressions from stars (e.g., winds,
supernovae, HII regions) and supersonic turbulence generated on
larger scales also form self-gravitating clouds (e.g., Hartmann, et
al. 2001), just as they form diffuse clouds. Because self-gravity is
involved at some point in this formation process, whether at the
beginning for the spiral instabilities or at the end for the
collect-and-collapse scenarios, and because the induced motions
which start the latter are supersonic, the time scale for
self-gravitating cloud formation is relatively short. That is, it
operates in about a crossing time, which is also about the dynamical
time, $t_{dyn}=\left(G\rho\right)^{-1/2}$.

For the topic of the present section, there are two important
points: cloud formation itself does not involve or require magnetic
diffusion, and self-gravitating cloud formation begins in an ambient
ISM that is close to magnetically critical on a large scale. The
first point illustrates again the relatively minor role of magnetic
diffusion in star formation, limited, as it appears to be, to the
final stages. The second point is directly related to the proposed
rapid evolution of molecular cloud cores to supercritical collapse.
The steps leading to this collapse are considered here.

The galactic dynamo pumps energy from shear and turbulence into the
ambient magnetic field until the field pressure is comparable to the
other energy densities that give the gas layer its thickness. Higher
fields lose magnetic flux from the disk by the Parker (1966)
instability. At the same time, galactic evolution with its cycle of
self-gravitating cloud formation, star formation, and supernovae
tends to pin the Toomre instability parameter $Q$ at about unity
(Goldreich \& Lynden-Bell 1965). In that case, the self-gravitating
energy density in the ISM is comparable to the other energy
densities, and the disk thickness is about the ambient Jeans length.
For these reasons, the magnetic energy density is comparable to the
self-gravitating energy density on kpc scales in the main disks of
spiral galaxies. Locally, both have a value of about 0.5 eV
cm$^{-3}$. In this sense, the ambient ISM always has a near-critical
magnetic field.

If we imagine a box with a height equal to the gas layer thickness
(including HI) and a length and width equal to the inverse Jeans
wavenumber parallel to the galactic plane, then the box is nearly
cubical with all dimensions $\sim a^2/\left(\pi G\Sigma\right)$ for
velocity dispersion $a$ and mass column density $\Sigma$. This is
the basic unit of self-gravitating cloud formation on a galactic
scale: the basic unit for swing amplified and magneto-Jeans
instabilities, and the basic unit for gas before a stellar spiral
density wave shocks it into a filamentary dust lane. It might also
be the outer scale for turbulence driven by gaseous self-gravity
(Elmegreen, Elmegreen \& Leitner 2003; Kim \& Ostriker 2007). The
mass of the basic unit for local conditions is $10^7$ M$_\odot$
(much larger than a giant molecular cloud). The corresponding first
step of self-gravitating cloud formation has been called either a
supercloud (Elmegreen \& Elmegreen 1983, 1987) or a giant molecular
association (Rand \& Kulkarni 1990) depending on the molecular
fraction, which, in turn, depends on metallicity and pressure
(Elmegreen 1993; Honma, Sofue \& Arimoto 1995, Wong \& Blitz 2002)
and is unrelated to the cloud formation process.

We would like to know what happens to the state of magnetic
criticality as this basic, nearly-spherical, unit is distorted into
various shapes by large-scale processes, and as the gas inside the
unit recollects following these distortions into giant molecular
clouds and their nearly-spherical cluster-forming cores.  We show
that every nearly-spherical, self-gravitating condensation that
forms inside the basic unit will also be close to the critical field
limit, regardless of intermediate steps, and that this preservation
of criticality works quickly, on the dynamical time scale. In
section \ref{sect:dest} we discuss second-generation cloud core
formation in the filamentary debris of first-generation GMCs; such
filament streaming makes the cores more supercritical than the
debris.

First consider the most fundamental definition of magnetic
criticality, where the gradient in the field energy density, $\nabla
B^2/4\pi$, equals the self-gravitating force density, $g\rho$, for
field strength $B$, gravitational acceleration $g$ and density
$\rho$.  Along a flux tube of halfwidth $R$, the first is $\sim
B^2/4\pi R$ and the second is $2\pi G\rho^2R$. Their equality gives
the critical field strength $B=8^{1/2}\pi G^{1/2} \rho R$.
Similarly, for an infinite disk with a perpendicular mass column
density $\Sigma$, the critical field is $B=2\pi G^{1/2}\Sigma$
(Nakano \& Nakamura 1978). The coefficients differ by only a factor
of $2^{1/2}$. Here we write the critical field strength as
$B=X\Sigma$ for constant $X$. For an equilibrium 3D configuration,
the density and column density vary with position so either the
central $B/\left(G^{1/2}\Sigma\right)$ ratio or the central mass $M$
to magnetic flux flux $\Phi$ ratio are considered, or the total mass
to magnetic flux ratio. Tomisaka, Ikeuchi \& Nakamura (1988) find
for the central value $G^{1/2}\Sigma/B=G^{1/2}M/\Phi=0.17$, while
Mouschovias \& Spitzer (1976) find for the total cloud value
$G^{1/2}M/\Phi=0.13$. For generality, we write the critical mass to
flux ratio as a constant $M/\Phi=Y$.

The mass to flux ratio is an indicator of stability only for
spheroidal clouds, which are bounded in 3 dimensions. Suppose a
large round cloud threaded with field lines is critical with both
$B=X\Sigma$ and $M/\Phi=Y$; these expressions have the same meaning
for a round cloud. A thin tube of flux inside this cloud has about
the same $B$ and $\rho$ but a smaller radius $r<<R$, and so it is
magnetically sub-critical ($B>>X\rho r$) by the first definition.
However the flux in this tube is smaller than the flux in the whole
cloud by the ratio $(r/R)^2$, and the mass in the tube is smaller
than the whole cloud by the same ratio, so the mass to flux ratio of
the tube is the same as in the whole cloud: $M/\Phi=Y$. Turbulence
compression perpendicular to the magnetic field can form a small
tube of flux like this. To be specific, suppose compression changes
the cross-field dimension in part of the cloud by the geometric-mean
factor $C$ (i.e., $C=R/r>1$). Then without magnetic diffusion, both
the flux and the mass in this compressed region are the same as they
were before, rendering $M/\Phi$ unchanged at the value $Y$. At the
same time, $B$ increases by $C^2$ and $\Sigma$ by $C$, so $B/\Sigma$
goes up by the factor $C$ to the value $XC$. Such a compressed
region is stable in the transverse direction because it would expand
back without the confining ram pressure of the turbulent flow around
it. Thus the $B/\Sigma$ condition, which indicates stability in this
example ($XC>X$), is more fundamental for diverse geometries than
the $M/\Phi$ condition.

Now consider what happens to the state of criticality as gas flows
along the field lines in the compressed filament, collecting into
$N$ cores of height $H$ spaced out along its total length $L$.  The
mass in each is the initial mass of the filament divided by $N$, so
$M/\Phi$ decreases from $Y$ to the sub-critical value $Y/N$. The
transverse column density in each increases during this collection
by the factor $L/NH$, so $B/\Sigma$ becomes $XCNH/L$. If both the
cores and the original cloud are somewhat spherical, then $C\sim
L/H$ and the first condition also becomes subcritical, by the same
factor $N$: $B/\Sigma\sim XN$. Thus the cores are stable by the same
degree for both conditions. For them to form stars, magnetic
diffusion has to reduce the flux by a factor $N$. {\it The
observation of protostars or dense mm-wave continuum sources strung
out along filaments implies that significant magnetic diffusion has
already occurred.} During this diffusion, both criticality
conditions move together from sub-critical to super-critical. (We
note that turbulence simulations without magnetic fields also get
beaded filaments, but this field-free case is unrealistic. The same
result in the magnetically critical case requires diffusion.)

Compression parallel to the field should also be considered. Suppose
part of the cloud of length $L<R$ is compressed along the field into
a layer of thickness $H$, so the density and transverse column
density (perpendicular to the field) go up by the factor $L/H$. The
field strength will be unchanged at first by the parallel motions so
$B/\Sigma$ decreases to the {\it super}-critical value $XH/L$. The
$M/\Phi$ ratio drops for the layer because, although the flux is
constant, only part of the total cloud mass is involved; this gives
$M/\Phi\sim YL/R$, a {\it sub}-critical value. However, the layer
will be heavily weighed down transverse to the field and it will
adjust, pulling in the field with it. Because $B$ increases as the
inverse of the transverse area and $\Sigma$ increases only as the
inverse of the transverse length, the ratio $B/\Sigma$ goes back up
and eventually becomes $X$, stabilizing the collapse. This stable
point occurs when the new transverse radius is $r\sim RH/L$.  The
transverse collapse preserves both mass and flux in the layer, so
$M/\Phi$ stays with its sub-critical value $YL/R$. Thus the new core
is critical by the first condition and sub-critical by the second
condition. However, prior to the initial compression, this part of
the cloud had the same low $M/\Phi$ ratio as it did after the
compression and collapse: the mass was down by the factor $L/R$ at
both times. Thus both $B/\Sigma$ and $M/\Phi$ are unchanged after
the adjustment for this part of the cloud. Moreover the final
condensed object will be spheroidal if the transverse size, $r\sim
RH/L$ equals the parallel height, $H$, and this requires $L\sim R$.
Thus parallel compressions and equilibrium adjustments in spherical
clouds leading to spherical cores will leave the state of
criticality unchanged if there is no magnetic diffusion.

In general, there will be compressions from turbulence and external
pressures both transverse and parallel to the field, and at oblique
angles, and the compressed gas will re-adjust by self-gravitational
forces forming dense cores. Transverse components of compression
tend to make cores with a decreased state of criticality (for $N>1$
in the above example) and these cores will survive only during the
active compression unless there is significant flux loss at this
time. Parallel components of compression tend to preserve the state
of criticality even without flux loss.  If the characteristic length
for collapse along a transversely-compressed filament is much longer
than its width (as suggested by theory -- Fiege \& Pudritz 2000),
and comparable to its length, then only $N=1$ core will form in the
transverse-compressed filament and both directions of compression
produce cores with the same state of criticality as the initial
cloud. In this case, cloud formation by self-gravitational
readjustment of compressed and distorted basic ISM units will
produce whole GMCs at about the same state of magnetic criticality
as the ambient medium. Similarly, the round cores of GMCs, however
they form, will begin their lives close to the state of magnetic
criticality of the surrounding cloud. As the ambient ISM is
approximately magnetically critical, so the clouds and
cluster-forming cores will be too, before any diffusion begins.
V\'azquez-Semadeni et al. (2005) also noted that spheroidal
supercritical cores require spheroidal supercritical clouds in the
absence of magnetic diffusion.

We note that conservation of magnetic criticality from a large
spheroidal cloud to a small spheroidal cloud implies significant
movement of gas along the field lines in the absence of magnetic
diffusion. We are suggesting this is the case. Such parallel motion
does not happen all at once, however, but in steps as the GMC and
GMC core form in a hierarchical fashion from self-gravity and
turbulence compression. The parallel motion is also accompanied by
perpendicular motion to preserve cloud roundedness, but the latter
moves the field with it, causing the observed hour-glass shape with
a pinch at the middle.  Eventually magnetic diffusion smooths out
the field, but this smoothing is not needed for GMC core-formation,
which operates quickly. A typical GMC is 1\% the mass of a
supercloud ($10^5$ M$_\odot$ compared to $10^7$ M$_\odot$), and 10\%
of the size ($100\times20\times20$ pc$^3$ compared to
$1000\times200\times200$ pc$^3$). This means the GMC is a
contraction of the inner $100^{-1/3}\sim0.2$ of the supercloud
length by a factor of only $\sim2$. Similarly, a cluster-forming
core might be $\sim10$\% of the GMC mass and 20\% of the size, which
means that $10^{-1/3}=0.46$ of the GMC length is involved in core
formation with a contraction that is another factor of $\sim2$. Most
of the supercloud mass stays in a low density envelope that has
little star formation, and most of the GMC mass stays in another,
denser envelope where star formation is also relatively slow.  The
two factors of 2 illustrate the modest amount of motion along field
lines and the mild perpendicular re-adjustment to this motion.  The
smallness of these factors helps explain the quickness of GMC and
GMC core formations.

At this point in the evolution of a cloud, it takes only a little
magnetic diffusion for the core to become supercritical and begin a
collapse phase. Whereas the usual diffusion time is $t_B=R/\Delta v$
for ion-neutral drift velocity $\Delta v$ and cloud radius $R$, the
diffusion time to reach a supercritical state from an initial state
with $B=aX\Sigma$ and $a$ slightly larger than 1 is only
$t_B\left(1-1/a\right)<<t_B$. This is because $\Sigma$ has to
increase by $1/a$ to become supercritical, and for transverse
motions $\Sigma\propto 1/R$. Thus the core gas has to diffuse only
from an initial radius $R$ to a slightly smaller radius $R/a$ for
the core to go supercritical, and the time scale for this is
$\left(R-R/a\right)/\Delta v=t_B\left(1-1/a\right)$.  A more
detailed discussion is in Ciolek \& Basu (2001). They get
$t_{core}\sim t_B\left(1-M/Y\Phi\right)$ for marginally subcritical
cores, and this is about the same result.

\section{Time Scale for Star Formation}
\label{sect:time}

\subsection{The Big Scale}
\label{sect:bigfast}

One of the most revealing diagnostics of the star formation process
is the time scale. If it is long compared to the dynamical time,
then mechanisms for delay and prevention of star formation in
strongly self-gravitating gas have to dominate the process. If the
ratio of times is of order unity, then the gas collects into stars
as fast as physically possible.

Modern observation suggest that star formation is much faster than
we imagined three decades ago, when a time scale of $\sim10^8-10^9$
yrs came from the ratio of Galactic molecular mass to star formation
rate (Zuckerman \& Evans 1974; Scoville \& Solomon 1975) and from
the high molecular fraction of the inner Milky Way (Solomon, Sanders
\& Scoville 1979). The GMC evolution times dropped to 30 Myr when
the interarm regions were found to be mostly free of GMCs (Bash,
Green \& Peters 1977), and to 5-10 My when the ages of newly exposed
clusters were first observed (Leisawitz, Bash \& Thaddeus 1989). Now
the formation times of dense clusters can be observed directly,
suggesting that the main star formation activity is often over in
$\sim3$ Myr. Inside the denser cores where individual stars form, it
can be as short as several $\times10^5$ years (e.g., Onishi et al.
2002). In the LMC, GMC lifetimes may be a little longer $\sim10-30$
Myr for reasons that are unexplained (Fukui 2007; Kawamura et al.
2007).

All of these time scales are still correct in the sense that the
basic observations have not changed. The interpretation of what they
mean has changed, however. The high molecular fraction in the inner
Milky Way implies that gas spends a high fraction of its time in
molecular form without forming stars. This means some combination of
four things: (1) star formation is inefficient in CO clouds; (2)
molecular envelopes contain most of the CO mass but evolve more
slowly than the dense molecular cores where stars form; (3)
molecular clouds get dispersed in CO-rich pieces after star
formation, and (4) some molecules are in diffuse clouds that do not
form stars (Polk et al. 1988).

These four points are familiar: self-gravitating clouds are
assembled by gravitational collapse in spiral density wave shocks,
turbulent shocks, and explosive shells, and they form by
gravitational instabilities in the ambient medium as part of
swing-amplified spiral growth. These are all dynamical processes
that operate as quickly as possible in the ISM, such as the crossing
time over the scale height, $H=a^2/\pi G\Sigma$, or the self-gravity
rates $\left(G\rho\right)^{1/2}$ and $\pi G\Sigma/a$ for average
midplane density $\rho$, disk column density $\Sigma$ and velocity
dispersion $a$. The clouds then evolve to high density on the
internal dynamical time scale which decreases as the density
increases. Molecules form in clumps as soon as the cloud can shield
itself, which is at a relatively low average density for a massive
cloud (e.g., Pelupessy, Papadopoulos, \& van der Werf 2006; Glover
\& Mac Low 2007), and stars form slightly later. The main point here
is that even though every step operates at close to the local
dynamical rate, the star formation step operates at the highest
density where the dynamical rate is greatest. Cloud core disruption
is also at the high core rate, via shocks from winds and HII
regions. Thus self-gravitating clouds spend a longer time forming
than getting dispersed even though every step evolves as quickly as
possible: the relevant density is lower when they form than when
they get dispersed.

Cloud destruction after star formation involves mostly the dense
core. Part of the GMC envelope will get compressed during core
disruption and form new stars as a result, and part will get pushed
away with only scattered star formation before settling into new
cores.  In either case, a large fraction of the GMC molecules
outlasts the first generation of star formation in the core, which
typically involves only $\sim10$\% of the GMC mass. If the ratio of
the core density to the average density in a GMC is $\sim100$, then
the ratio of dynamical times is $\sim10$, and the envelope molecules
last $\sim10$ times longer than the star formation event. The
fraction of GMCs that are active is not 10\%, however. The timing
factor of 10 has to be divided by the number of dense-core locations
and generations per GMC. Considering that the Orion cloud formed
$\sim4$ generations and other local clouds form a similar number of
subgroups, and that star formation usually persists at a relatively
low level on the periphery of OB associations even after the dense
core phase is over, the inactive GMC fraction is very low. Thus the
fraction of the ISM in the form of GMCs can be high, as it is in the
inner Milky Way, even though each stage prior to star formation
evolves at the local dynamical rate. The star formation rate is low,
only $\sim1$\% of the dynamical rate for the average GMC, because
$\sim90$\% of the GMC mass has a long dynamical time, and
$\sim80-90$\% of a GMC core gas gets dispersed during star formation
(i.e., cluster formation in dense cores is only $\sim10-20$\%
efficient; e.g., Tachihara et al. 2002; Lada \& Lada 2003; Brooke et
al. 2007; J{\o}rgensen et al. 2007).

It is important to make a distinction here between GMC destruction,
where the GMC is converted back into atomic form, and GMC dispersal,
where the GMC is moved and broken apart. Both lead to the end of
star formation in any one location, but the relevant time scales and
processes differ and their contributions to the total molecular
fraction differ.  GMC destruction requires ionization and heating,
so the cloud is disassembled molecule by molecule. Ionization can
destroy part of a GMC, but ionization is usually accompanied by
compression and motion, so the cloud moves away in pieces before it
is completely destroyed. Whitworth (1979) estimated that the ionized
mass from an embedded OB cluster is $2.3\times10^4\left(T_i/5\;{\rm
Myr}\right)^3\left(\epsilon/0.04\right)^{4/3}\left(n/10^3\;{\rm
cm}^{-3}\right)^{-1/3}$ $M_\odot$ for ionization time $T_i$, star
formation efficiency $\epsilon$ and cloud density $n$. With shorter
O-star lifetimes than he assumed, $T_i\sim3$ Myr, and slightly lower
$\epsilon$ for whole OB associations (e.g., 1\% per generation for
average GMCs according to Williams \& McKee 1997), the ionized mass
is only $10-20$\% of a GMC mass. Thus there is usually a large mass
from the GMC envelope left over after cluster formation in the core,
and this mass is available for more star formation in a slightly
different location after another dynamical time (Sect.
\ref{sect:dest}).

It is also important to distinguish between timescales for GMC
destruction, dispersal, and consumption.  The latter time is the
total GMC mass divided by the galactic star formation rate.  We
suggest that the dispersal time is the fastest of these and is
comparable to the dynamical time because star formation begins and
ends quickly in a GMC. The destruction time is longer because each
GMC may go through several stages of star formation following
disruption in active cores. The consumption time is longest because
the efficiency of star formation in each event is low and relatively
little gas gets used up. GMCs exist in one place for a dispersal
time and they exist as entities for a destruction time. There is no
physical meaning to the consumption time as far as an individual GMC
is concerned.

The fourth point mentioned above, that there are molecular diffuse
clouds in the inner galaxy, follows from the fact that molecular
self-shielding is independent of cloud self-gravity, depending more
on the product of density and column density in the shielding layer
than on any property of the cloud interior. High pressure regions
have higher diffuse cloud densities in thermal equilibrium with the
radiation field, and so require lower column densities for
self-shielding. Thus the diffuse molecular mass can be high in high
pressure regions, which includes the inner parts of galaxies and
starburst galaxies (Elmegreen 1993).  The observed galactic
gradients in molecular fraction are partially the result of this
pressure gradient combined with a metallicity gradient (Honma, Sofue
\& Arimoto 1995; Wong \& Blitz 2002). In M64, 25\% of the CO
molecular mass is diffuse (Rosolowsky \& Blitz 2005).

These points illustrate how the molecular mass can be high and the
star formation rate low.  The star formation rate is the efficiency
per cloud multiplied by the cloud formation rate, which equals the
destruction rate when the total CO mass is constant. The efficiency
is very low in GMCs, a few percent (Williams \& McKee 1997), so the
consumption time, which is the ratio of the cloud mass to the star
formation rate, is long. The efficiency is low for GMCs because only
a small fraction of the GMC mass is involved with active star
formation. These active regions are, for example, HCN cores (Sect
\ref{sect:hcn}). In such cores the total efficiency is higher than
it is in a GMC by the cloud to core mass ratio. The formation and
destruction rates of HCN cores could by dynamical as well, and much
faster than the CO formation and destruction rates because of the
higher density. As for CO, the HCN core lifetime is the ratio of the
HCN mass to the star formation rate divided by the efficiency in the
HCN core. This efficiency is higher than in the CO cloud, but it
still quite small because the real action happens in even denser
sub-cores (e.g., the CS cores) which have an even lower total mass
and a higher local efficiency (e.g., Shirley et al. 2003).
Eventually a high enough density should be reached where the total
galactic mass divided by the local dynamical time is within a factor
of a few times the total star formation rate. There, the efficiency
will be high, 30\% to 50\%, and the final contraction to a unique
star take place.  The correspondence between decreasing scale and
increasing efficiency is expected for hierarchically structured
clouds.

Dense, high-efficiency regions of star formation have probably been
observed. The mm-wave continuum cores and other dense cores that
have a Kroupa (2001) mass function show a shift in the turnover mass
at some value that is higher than the stellar IMF turnover by a
factor of about 3 (Motte et al. 1998; Testi \& Sargent 1998;
Johnstone et al. 2000, 2001; Motte et al. 2001;  Beuther \& Schilke
2004; Stanke et al. 2006; Alves, Lombardi \& Lada 2007). If these
dense cores form individual and binary stars, then their
efficiencies are the inverse of this factor.  This final stage has
an extremely fast dynamical time compared to that in the lower
density gas. The contraction can be significantly retarded by
magnetic forces without affecting our proposal that cluster-forming
cores (which have lower average density) evolve in a dynamical time.
The bottleneck in the star formation processes is at the lower
densities, where the evolution is slow because the dynamical time is
long.

The historical decrease in molecular cloud lifetime that was
mentioned at the beginning of this section has a simple explanation.
Generally, as the scale of the region observed has decreased with
improved instruments and finer surveys, the lifetimes of the clouds
that are seen have dropped. Gas structure has a wide range of scales
and no characteristic scale, as shown by the power law power spectra
of HI (Dickey et al. 2001) and CO (St\"utzki et al. 1998). Surveys
with particular sampling sizes tend to highlight structures with a
narrow range of scales, from several times the sampling size
(Verschuur 1993) to several tens of the sampling size, as which
point the largest structures tend to be ignored in favor of the
clumps inside these structures. That is, clusters of clumps are not
found by clump-finding algorithms. The algorithms only find
continuous regions, and these are always close to the resolution
limit of the telescope or survey. Because all structures evolve on
their local dynamical time, higher resolution surveys that find
smaller clouds also observe smaller lifetimes.

For example, giant spiral arm features (``beads on a string'') that
produce star complexes 300 pc in diameter remain active for
$\sim30-50$ My (Efremov 1995), GMCs that produce OB associations
(which are generally inside star complexes in a hierarchical sense;
e.g., Battinelli, Efremov \& Magnier 1996) last $10-20$ Myr, while
GMC cores that produce clusters last only $\sim3$ Myr. Generally
there are several generations of small-scale star formation inside
each large-scale region (Efremov \& Elmegreen 1998). Thus, star
formation is hierarchical in time as well as space. This double
hierarchy can be misleading because observations always contain
selection effects. OB associations and the $10^5$ M$_\odot$ GMCs
that make them are not a characteristic scale for star formation
even though they all have about the same size in normal galaxy
disks: $\sim80$ pc (Efremov 1995). They are selected for that size
by the selection of a star-formation time scale through the survey
requirement that O stars are still present. Once a survey is about
OB associations, the size of the region is fixed by the dynamical
time scale. This size will be larger in higher pressure regions
because of the way size $R$ and dynamical time scale with pressure
$P$: $R^2/\left(t_{dyn}^4P\right)\sim G/2.5$ (Elmegreen 1989). The
associated stellar and cloudy masses will be larger in high pressure
regions too: $M^2/\left(P^3t_{dyn}^8\right)\sim1.6G$. The same can
be said for other stellar clusters selected by age, such as T-Tauri
associations with smaller $t_{dyn}$ and star complexes with larger
$t_{dyn}$, as identified by Cepheid variables and Red supergiants
(Efremov 1978).  For these reasons, more recent surveys with smaller
scale resolutions get shorter cloud lifetimes.

\subsection{The Small Scale}
\label{sect:smallfast}

An important consideration is how far down in scale the dynamical
evolution goes, and whether it slows when star formation begins at
the bottom. Elmegreen (2000) compiled evidence that a dynamical
cascade persists down to the scale of embedded star clusters without
significant delay on the small scale. The time scale for each level
was said to be $\sim1-2$ crossing times where the crossing time was
taken to be $R/V\sim1.09/\left(G\rho\right)^{1/2}\sim t_{dyn}$ for a
uniform virialized cloud. For molecular density $n$,
$t_{dyn}\equiv\left(G\rho\right)^{-1/2}=61n^{-1/2}$ Myr. Star
formation cannot be as fast as a single crossing time or shorter
than a crossing time (unless there is an implosion -- Lintott et al.
2005) because turbulent and magnetic energy has to dissipate; thus
1-2 crossing times seemed to be a reasonable match to the
observations. Tan, Krumholz \& McKee (2006) suggested there is a
delay for clusters, amounting to $\sim4-5$ crossing times. They
suggest the longer time requires near-equilibrium cloud support.
Here we review the issues raised in these papers.

The discussion in Elmegreen (2000) had four points: (1) hierarchical
structure in young stars often mimics hierarchical structure in
molecular clouds, implying these stars had less than a crossing time
to mix, (2) embedded clusters ages are relatively small, (3) age
differences between neighboring clusters are relatively small, and
(4) a high fraction of dense cores contain young star formation. The
case for short star formation times was independently made by
Ballesteros-Paredes, Hartmann, \& V\'azquez-Semadini (1999), based
on the short duration of star formation in Taurus.  A recent review
of short time scales is in Ballesteros-Paredes \& Hartmann (2007).

The discussion in Tan, Krumholz, \& McKee (2006) had six points: (1)
CS clumps are nearly round, (2) clusters are generally smooth, (3)
protostellar wind momentum, which is proportional to the star
formation rate, is small, (4) cluster age spreads are relatively
large, (5) a dynamical ejection event in the Orion Nebula cluster
occurred a relatively long time ago (2.5 Myr), and (6) stellar mass
segregation requires a relatively long time.  The third method using
wind momentum is highly inaccurate, as these authors admit, so we
will not use it here to determine a cluster formation time to within
the desired factor of 3.

The other points are reconsidered here. We begin with the usual
caution that star forming regions have a range of densities so the
crossing time does not have a single value. The dynamical time will
also be longer for the lower density regions, so the total age
spread for stars should be larger than the age spread in a cluster
core. As clouds contract and the density increases, the crossing
time decreases, so prior star formation in the same cloud will have
a longer time scale than current star formation in the core (we show
a model of this in Sect. \ref{sect:huff}). Thus, fairly old stars
should always be present in an active region even if the current
level of activity is short-lived. Palla \& Stahler (2000) have
termed this evolution accelerated star formation: the star formation
rate in a region increases with time until the final cloud core is
disrupted. Huff \& Stahler (2006) found an extended age distribution
in the Orion Nebula cluster and modeled it with continuous star
formation during monotonic cloud collapse. There was no equilibrium
or energy feedback in their model and yet the evolutionary time
scale matched the stellar ages. For these reasons, we do not
consider the observation of relatively old stars or relatively old
ejection events to be an indication that cloud cores are stable.
This point is relevant to some of the discussion in Tan, Krumholz,
\& McKee (2006), and also counters most of the discussion in Tassis
\& Mouschovias (2004).

There remain two lifetime indicators that are based on morphology
alone: clump shapes and cluster substructure. Tan, Krumholz \& McKee
(2006) note that CS cores forming high mass stars are circular to
within $\sim26$\%. Equilibrium clouds are also round so they
conclude the CS clouds are in equilibrium. However, numerical
simulations form roundish objects that are not in equilibrium
(Ballesteros-Paredes \& Mac Low 2002; Gammie et al. 2003; Li, et al.
2004), and other cores are more irregular than the CS observations
suggest (Myers et al. 1991; Bacmann et al. 2000; Steinacker et al.
2005).  The CS sources cited by Tan, Krumholz \& McKee were observed
by Shirley et al. (2003) with a 24.5 arcsec beam at an average
distance of 5.3 kpc, making the average beam half-size 0.31 pc. The
average deconvolved radius of a CS core was calculated to be 0.37
pc, which is about the same. Thus the sources are barely resolved.
The major and minor axes used by Shirley et al. to obtain the
average ellipticity measurement of 1.26 did not consider deconvolved
beams, however, so the intrinsic ellipticity ratios are higher than
the observed ratios. Higher resolution observations may eventually
show irregular substructures like those commonly seen at lower
densities; then the CS cores would appear to be more rapidly
evolving.

Cluster substructure consists of stellar hierarchies and filaments
that mimic cloud hierarchies and filaments (Gomez et al. 1993; Testi
et al. 2000; Heydari-Malayeri et al. 2001; Nanda Kumar, Kamath, \&
Davis 2004; Smith et al. 2005; Gutermuth et al. 2005; Stanke, et al.
2006; see review in Allen et al. 2006). The clusters have to be
fairly young to show it, and even then it appears most prominently
in the youngest protostars (e.g., Dahm \& Simon 2005). By the time a
cluster core is ready for disruption, which may be 1.5 crossing
times after its most active phase began, the oldest stars will have
moved around enough to mix their birth sites and only the
pre-stellar cores and youngest protostars will still show gas-like
morphologies.

The discussion about this in Section 2.2 of Tan, Krumholz, \& McKee
(2006) has a different conclusion. They consider a cluster mass $M$
forming in a total time $t_{form}$, and a total mass in subclusters,
$M_{sub}$, which each form and disperse in a time $t_{dyn}$.  For
unbound subclusters, the steady state gives
$t_{form}=\left(M/M_{sub}\right)t_{dyn}$. For bound subclusters,
they consider the individual lifetimes to be the time for a
subcluster to sink to the cluster center by dynamical relaxation,
which is $\sim0.17\Lambda\ln\Lambda$ dynamical times for $\Lambda$
equal to the ratio of the cluster mass to the mass of an individual
substructure. For IC 348, which has 8 subclusters with 10-20 stars
each out of the total of 345 stars, the unbound case gives
$t_{form}\sim5t_{dyn}$ and the bound case in their analysis gives
$t_{form}\sim21t_{dyn}$. We note that $t_{dyn}$ for the unbound case
was assumed to be the dynamical time for the substructure, which is
less than the dynamical time for the larger-scale core according to
the usual time-size-velocity scaling relation. For example, if the
substructure is 1/4 of the core size and the dynamical time scales
with the square root of size as in the Larson (1981) law, then
$t_{form}$ would be only $2.5t_{dyn}$ in the unbound case. However,
the basic model should be questioned. Bound and unbound
substructures should interact with each other more frequently than
they evolve on their own.  A substructure with any reasonable size
cannot cross from one side of the cloud core to the other without
mixing with another substructure. Colliding loose substructures
either merge or destroy each other if their collision speed is less
than a few times their internal dispersion (e.g., Aarseth \& Lecar
1975). Their lives are much shorter than either their own
dissolution time or their sinking time from dynamical friction. For
example, the filamentary structures seen in young mm-wave continuum
sources and the hierarchical structure seen in young stars and
protostars have characteristic outer scales that are comparable to
the scales of the cluster cores. If stars are born with the same
pervasive hierarchy as the gas, which extends over all available
scales, then each subcluster can hardly move without interacting
with another one. Only the smallest and densest might last for a
full crossing time. Thus clusters substructure should be evanescent
with average individual lifetimes much less than a core crossing
time.  This short time accounts for the low fraction of clusters
with substructure even when the star formation time is only 1-2
crossing times.

Krumholz \& Tan (2007) continue the discussion of relatively long
time scales by comparing the star formation efficiency per unit free
fall time, $\epsilon_{ff}$, for a variety of molecular tracers. The
free fall time is $t_{ff}=\left(3\pi/32G\rho\right)^{1/2}$, which is
$\left(3\pi/32\right)^{1/2}=0.54$ times the crossing time, so a star
formation efficiency of 1\% in a free fall time corresponds to a
star formation efficiency of 1.8\% in a crossing time. They suggest
the average $\epsilon_{ff}$ is only a few percent, independent of
density, and so cluster formation with a final $\sim10$ percent
efficiency requires $\sim5$ free fall times. There are several
points to make here. First, the observed total efficiency for whole
OB associations is only a few percent per generation (Williams \&
McKee 1997), so $\sim1.5$ free fall times in a GMC ($=0.75$ crossing
times) is a reasonable result for these large scales, and it
requires the GMC evolution time to be short as suggested here.
Second, a slight increase in $\epsilon_{ff}$ with $\rho$ in Figure 5
of Krumholz \& Tan (2007) was not mentioned but it might be expected
for several reasons. In the Krumholz \& McKee (2005) model,
$\epsilon_{ff}$ scales with the inverse cube root of Mach number
(their eq. 30), and the Mach number could decrease for higher
density regions. If we consider the Larson (1981) correlations as
Krumholz \& McKee do, in which the linewidth scales approximately as
$\rho^{-1/2}$, then the inverse cube root of Mach number scales with
$\rho^{1/6}$ for a constant temperature, and this is about the trend
in $\epsilon_{ff}$ with $\rho$ in Krumholz \& Tan (2007). On the
other hand, the linewidth-density relation in Plume et al. (1997)
goes the opposite way for dense gas, $\Delta v\propto \rho^{0.3},$
and then $\epsilon_{ff}$ would not increase with $\rho$ if
$\epsilon_{ff}\propto\Delta v^{-0.3}$ from the Krumholz \& McKee
theory. On a more general level, an increase in $\epsilon_{ff}$ with
$\rho$ should be expected regardless of the dynamics for
hierarchically structured clouds, because the mass fraction of dense
star-forming clumps always increases with the average density
(Elmegreen 2005).  This is what the observations in Krumholz \& Tan
show most directly, and it does not support or refute any particular
model of star formation.

The highest-density value for $\epsilon_{ff}$ in Krumholz \& Tan
(2007) comes from CS emission, and these authors suggest it is
overestimated by a factor of a few because of undersampling in the
CS surveys by Plume et al. (1997) and Shirley et al. (2003), who
observed only H$_2$O maser sources. Krumholz \& Tan assume
$L_{CS}>20$ L$_\odot$ from Plume et al., derive a conversion of
$M_{CS}/M_\odot=4.5\times10^4L_{CS}/L_\odot$, and get a CS mass
limit of $>9\times10^4$ M$_\odot$.  This is divided by a star
formation rate of 3 M$_\odot$ yr$^{-1}$ and by the free fall rate at
the beam-diluted average CS density of $1.8\times10^5$ cm$^{-3}$ to
get $\epsilon_{ff}<0.27$.  Shirley et al. (2003), however, state
that $L_{CS(5-4)}=20$ L$_\odot$ is the most likely value for the
Milky Way after considering various completeness corrections, so the
Krumhjolz \& Tan value of $\epsilon_{ff}$ may not be so high. Even
so, with a factor of 3 downward correction for $L_{CS}$, the
$\epsilon_{ff}-\rho$ correlation in Krumholz \& Tan is still present
because the plotted CS point is 10 times higher than the others at
the same density. More important to the present paper is the
observation in Plume et al. and Shirley et al. that the ratio of
bolometric luminosity to virial mass in CS gas is a factor of
several hundred higher than in CO gas, and also relatively constant
from region to region spanning a factor of 100 in gas mass. Thus the
CS gas is closer to the star formation stage than CO. The efficiency
should increase in this way with density as the observations zero-in
on the individual star-forming cores. This takes us back to the
fundamental property of hierarchical clouds mentioned above, that
the mass fraction of the densest cores increases with the average
density.

The most peculiar point in the Krumholz \& Tan diagram is the low
value for HCN, which has $\epsilon_{ff}\sim0.0058$. This is lower
than the other values at the same density by a factor of $\sim10$
and the corresponding long time for HCN evolution could raise
questions about the time scale for star formation. We discuss this
HCN value now.

\subsection{The Evolution Time in HCN Cores}
\label{sect:hcn}

Gao \& Solomon (2004a,b) and Wu, Evans, et al. (2005) derived the
star formation rate in HCN clouds using the associated IR
luminosity. They observed the proportion $L_{IR}\sim 900L_{HCN}$
$L_\odot \left({\rm K \;km\; s}^{-1} {\rm pc}^2\right)^{-1}$, and
then converted the $L_{IR}$ to a star formation rate with ${\dot M}=
2\times10^{-10} \left(L_{IR}/L_\odot\right)$ M$_{\odot}$ yr$^{-1}$
from Kennicutt (1998). They converted $L_{HCN}$ to a mass using the
virial theorem, $M_{dense}=\alpha L_{HCN}$ with $\alpha\sim10$
M$_\odot$ $\left({\rm K\; km\; s}^{-1} {\rm pc}^2\right)^{-1}$.  As
a result, the time scale is
\begin{equation}
{{M_{dense}}\over{{\dot
M}}}=\left({{M_{dense}}\over{L_{HCN}}}\right)\left({{L_{HCN}}\over
{L_{IR}}}\right)\left({{L_{IR}}\over{{\dot
M}}}\right)\sim5.5\times10^7\;{\rm
yrs},\label{eq:time}\end{equation}which is much longer than the
dynamical time at the high density of HCN. The virial theorem
conversion factor comes from the equations $M=5R\Delta v^2/G$ and
$L_{HCN}=T\left(8\ln2\right)^{1/2}\Delta v \pi R^2$ for Gaussian
dispersion of the emission line $\Delta v$ and source radius $R$.
These equations give $\alpha=2.1n(H_2)^{1/2}/T$ for density
$n(H_2)=3M/\left(4\pi R^3m(H_2)\right)$. HCN requires excitation at
$n=3\times10^4$ cm$^{-3}$ and Gao \& Solomon assume $T=35$K, which
gives $\alpha=10$ as above.

Gao \& Solomon (2004a,b) observed unresolved HCN emission from whole
galaxies, while Wu et al. (2005) observed individual star-forming
regions in the Milky Way. The $L_{IR}-L_{HCN}$ correlation was about
the same for each, and this is a bit surprising. In whole galaxies,
the IR comes from massive stars whether they are inside or outside
the dense neutral cores, whereas the HCN comes only from the cores.
Generally O-type stars disperse their cores and break out quickly,
long before they supernova. The time spent inside a core, which is
the star formation timescale of interest for this paper, can be
arbitrarily short for the same total $L_{IR}$, $M_{dense}$ and
${\dot M}$.  For example, an O-type star might spend its first 0.2
Myr inside an HCN core before disrupting it and then spend the
remainder of its 3 Myr lifetime outside HCN cores. Then the star
formation rate in a core is $3/0.2$ time the average rate, and the
HCN timescale per core is $0.2/3$ of 55 Myr, or 3.7 Myr.
Correspondingly, $0.2/3=6.7$\% of O-type stars would be in HCN cores
at any one time. Clearly for whole galaxies, the HCN timescale
derived above is not equivalent to the duration of star formation
inside an HCN core.

The situation for individual star-forming regions is different. All
of the points in Figure 5 of Krumholz \& Tan (2007) other than HCN
were for individual regions, as was the Wu et al. (2005)
contribution to the HCN point. In this case, the duration of star
formation in an HCN core equals the efficiency of star formation
there multiplied by $M_{dense}/{\dot M}$, and this product can be
much less than $M_{dense}/{\dot M}$ alone. The average efficiency of
star formation in a dense core is half the final efficiency if stars
form with a uniform rate, and the final efficiency is probably only
$\sim10$\%, considering that $\sim90$\% of embedded clusters become
unbound after the gas leaves (which requires a low final efficiency
and rapid gas clearing; Lada \& Lada 2003).  Thus the average
efficiency during the embedded lifetime may be only $\sim5$\%, in
which case the duration of star formation in HCN cores is
$0.05\times55=$ 2.75 Myr.  If star formation accelerates in a cloud
core (Palla \& Stahler 2000), then the average stellar mass fraction
during the lifetime of the core is less than half of the final mass
fraction, and the core duration would be less than 2.75 Myr.

The duration of star formation in HCN cores should be the same
everywhere because HCN excitation corresponds to a certain density
and the dynamical time at that density is fixed. Thus the agreement
between the Gao \& Solomon correlation and the Wu et al. correlation
begins with a fundamental timescale that is the same for each and
then multiplies this timescale upward by about the same amount for
each. The multiplication factor for the galactic scale is the
inverse of the fractional time that O-type stars spend in HCN cores.
For smaller scales, it is the inverse of the average star formation
efficiency in an HCN core.  These factors may not be exactly the
same, but for the 9 orders of magnitude in $L_{IR}$ separating the
Gao \& Solomon results from the Wu et al. results, differences in
the conversion factors amounting to a factor of $\sim3$ will not be
noticed.

The lifetime of the HCN gas (55 Myr) still has to be explained. This
lifetime is much longer than a crossing time and in this sense faces
the same problem as the CO-emitting gas in the Zuckerman \& Evans
(1974) discussion.  The solutions could be similar too.  First, the
HCN cores as a whole should evolve more slowly than their denser
subcores where the stars actually form, so the HCN stage prior to
star formation should be slower than the stage after star formation
in a sub-core becomes disruptive. Second, HCN should be only
partially converted into stars during cluster formation (the low
efficiency) and the residual should be pushed aside at high pressure
without decreasing its density much. Then it stays HCN but has a
geometry temporarily unsuitable for star formation until it
recollapses in a different position later. Third, some HCN is
probably not in the form of strongly self-gravitating cores where
individual stars form, but is diffuse intercore gas that is at a
high enough density and opacity to excite HCN. In this sense, HCN
alone, like CO alone, is not the star-forming gas but provides an
envelope or intercore matrix to the star-forming gas, which is much
denser. Of all these, the low efficiency in each HCN core, discussed
above, is probably the dominant cause of the relatively long
consumption time.

High resolution observations of HCN regions should show a low
fraction of the gas actively involved with star formation, as do
observations of CO clouds. However, it may be that nearly every HCN
region still contains some star formation, which is the case for CO
too.  The inactive HCN, like the inactive CO, should be peripheral
or intercore gas in the immediate vicinity of star formation, and it
should not be strongly self-gravitating by itself. Dense
self-gravitating cores in the same regions are the more likely
precursors of young stars, and their contribution to the HCN mass
fraction should be low, like the 5-10\% estimated above. There
should also be evidence for HCN mini-shells, comets, and other
disturbances at these high densities. These should resemble
structures observed on much larger scales in CO. Their presence
would indicate that cloud core dispersal maintains a high density
for a crossing time while second generations of stars might be
triggered into forming.

CS-emitting gas is closer to the density where star formation
becomes highly efficient, and the consumption timescale for CS is
correspondingly shorter than for HCN. Plume et al. (1997) estimate
this timescale is 13 Myr based on the total CS mass divided by the
Galactic star formation rate. This is still much longer than the CS
crossing time (0.12 Myr), so the CS abundance has the same problem
as HCN and CO with probably the same solution: it either has to
recycle after discrete events or form stars in a small fraction of
its mass (sub-cores) where the local efficiency is higher.

\subsection{The Huff \& Stahler Accelerated Star Formation Model with Feedback}
\label{sect:huff}

\begin{figure}\epsscale{.6}
\plotone{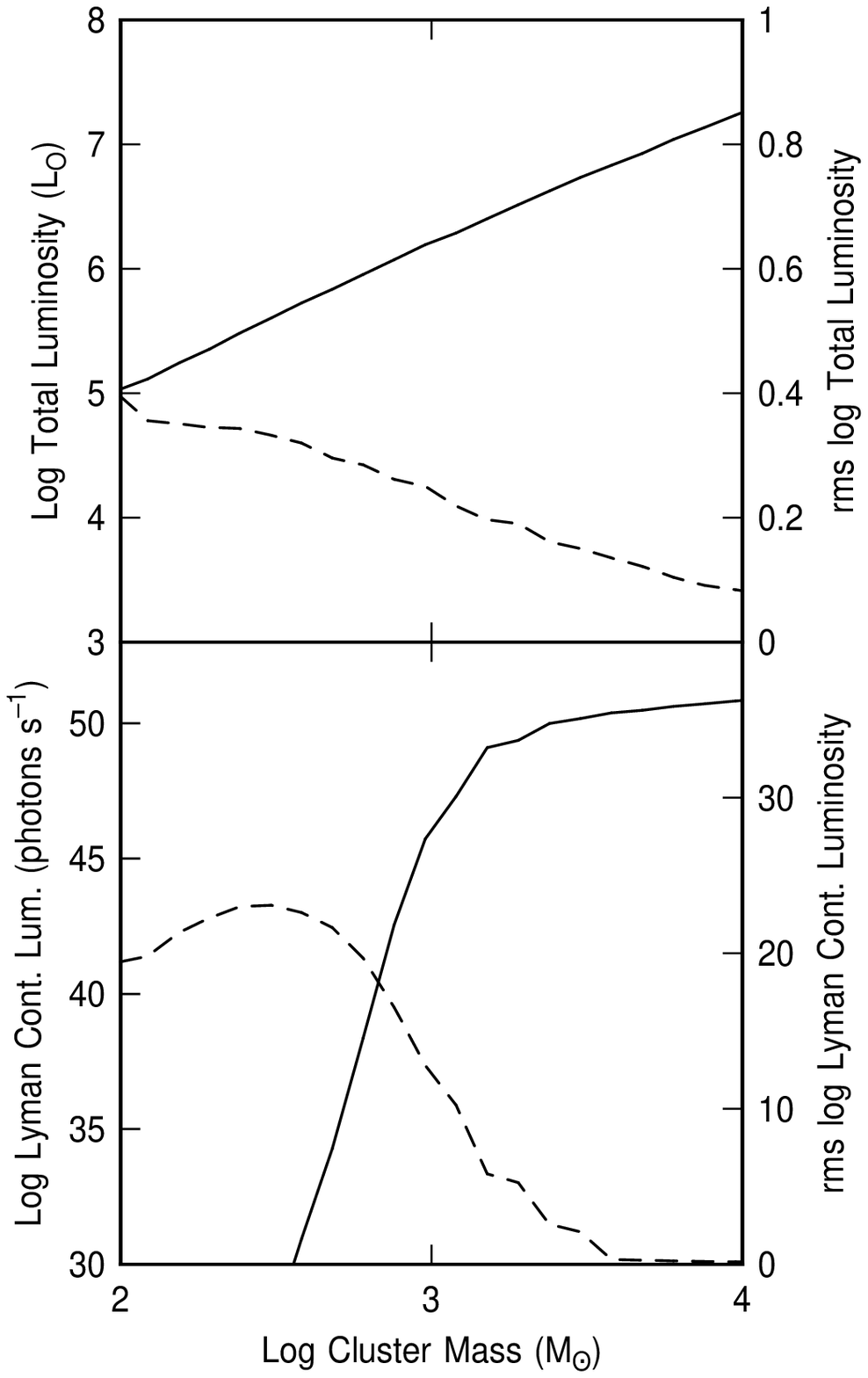}\caption{(bottom) The Lyman continuum luminosity in
photons per second (solid line) is shown versus the cluster mass,
using data in Vacca et al. (1996). 1000 cluster samples are made for
each mass, and each cluster is formed by randomly sampling an IMF.
The dashed line (right-hand axis) shows the rms deviation around the
average luminosity that comes from the stochastic sampling. The
Lyman continuum luminosity sharply increases for clusters more
massive than $\sim10^3$ M$_\odot$ because this is the mass where the
IMF is typically sampled out to the O-star range. (top) The total
luminosity of the cluster is shown versus the cluster mass, along
with the rms deviations from the sampling.
}\label{fig:vacca2}\end{figure}

Huff \& Stahler (2006) observed the history of star formation in the
Orion Nebula cluster and found accelerated star formation with some
stars older than the current crossing time. They suggested a model
of cloud evolution where energy dissipation removes turbulent
support and the cloud contracts, slowly at first, and then faster as
the dissipation and dynamical times get shorter. Their model has no
stellar energy input or feedback and still the cloud evolves
relatively slowly because it takes time to dissipate the turbulent
energy. Cloud contraction also generates more turbulent energy from
the change in $PdV$ for boundary pressure $P$ and volume $V$.  This
model is useful as a starting point to investigate star formation in
cores when feedback is added. To do this, we begin with the Huff \&
Stahler evolution equation for a singular isothermal sphere and add
an energy input term that is a function of the stellar mass
\begin{equation}
{{dH}\over{dt}}=-\eta{{M_{cloud}v^3}\over{2R}}+\Gamma
M_{star}^\gamma,
\end{equation}
where $M_{cloud}$ is constant, $M_{star}$ is the increasing stellar
mass, $v$ is the velocity dispersion, $R$ is the radius, and $H$ is
the enthalpy. The power $\gamma$ depends on the types of stars that
form. For a massive cluster where O-type stars form, $\gamma$ is
large because the luminosity, particularly beyond the Lyman
continuum, is a sensitive function of stellar mass. For a low mass
cluster or for a cluster where ionization is not important,
$\gamma\sim1$. We consider both cases here. Figure \ref{fig:vacca2}
shows the Lyman continuum luminosity (bottom) and the total
luminosity (top) as functions of the cluster mass. These curves were
obtained by randomly sampling the IMF until the desired cluster mass
was achieved. The IMF ranges from 0.01 M$_\odot$ to 150 M$_\odot$
with a flat slope below 0.5 M$_\odot$ and a slope of $-1.5$ above
that, for logarithmic mass bins (where the Salpeter slope is
$-1.35$). Each cluster mass used 1000 random trials and took the
resulting average luminosity. The rms variations around these
luminosities are shown by dashed lines using the right-hand axes.
The stellar luminosities and masses were obtained from Vacca,
Garmany, \& Shull (1996). The top panel shows that the total
luminosity increases approximately in proportion to the cluster mass
because the luminosity is heavily weighted by low mass stars. The
Lyman continuum luminosity is strongly dependent on cluster mass
($\gamma\sim40$ in places), with a sudden turn on of the Lyman
continuum flux at $M_{cluster}\sim10^3$ M$_\odot$, where O-type
stars first begin to appear. In the following models, we take
$\gamma=1$ and $2$ to illustrate the main points.

The kinetic energy in an expanding HII is approximately independent
of time and depends mostly on the initial thermal energy in the HII
region provided by the ionizing flux. This follows from the
expansion equation $R=R_0\left(1+7a_{II}t/4R_0\right)^{4/7}$, which
gives a velocity-squared proportional to $R^{-3/2}$, and from the
HII region mass, which is $M=(4\pi/3) m_H nR^3$ for
$(4\pi/3)nR^3=\left([4\pi/3]SR^3/\alpha\right)^{1/2}$. Here $S$ is
the Lyman continuum flux, $\alpha$ is the recombination coefficient
to the second level of hydrogen, $n$ is the HII region density, and
$R$ is the HII region radius.  The radius-dependence for mass and
velocity-squared cancel, leaving the expansion kinetic energy
constant at approximately its initial thermal value,
$0.5M_0a_{II}^2$, where $M_0$ is the Stromgren sphere mass at the
initial cloud density and $a_{II}$ is the thermal speed in the HII
region. The kinetic energy therefore depends on the square root of
the Lyman continuum flux. This justifies the use of this flux as a
crude measure of cloud-core destruction rates.

The stellar mass in our adaptation of the Huff \& Stahler model
comes from the volume and time integral over the instantaneous star
formation rate, which is taken to be
$\epsilon_{cr}\rho\left(G\rho\right)^{1/2}$ for constant efficiency
per crossing time $\epsilon_{cr}$. The volume integral gives
\begin{equation}
{{dM_{star}}\over{dt}}=\epsilon_{cr}\int_{R_{min}}^{R_0}
\rho\left(G\rho\right)^{1/2} 4\pi R^2dR=
\epsilon_{cr}\left({{GM_{cloud}^3}\over{4\pi}}\right)^{1/2}
{{\ln\left(R_0/R_{min}\right)}\over{R_0^{3/2}}}
\end{equation}
for a singular isothermal sphere with
$\rho(R)=\rho_0\left(R_0/R\right)^2$ (as assumed by Huff \& Stahler
2006). The collapse is assumed to decrease $R_0$ while keeping the
total cloud mass constant. The log term has assumed a minimum
radius, or inner core radius, $R_{min}$, to terminate the
singularity.  The enthalpy is $H=-GM_{cloud}^2/\left(12R\right)$
(Huff \& Stahler). The energy dissipation rate is assumed to be
$-\eta v^3/\left(2R\right)$ from Mac Low (1999), who determine
$\eta\sim0.4$. We take $\eta=0.3$ here to be slightly conservative
(small values lengthen the contraction and star formation time
scales relative to the crossing time). We also take
$\epsilon_{cr}=0.1$ and three values of $\Gamma$ for each case to
give a reasonable range for the final mass of stars.  The equations
are integrated numerically over time.

\begin{figure}\epsscale{1}
\plotone{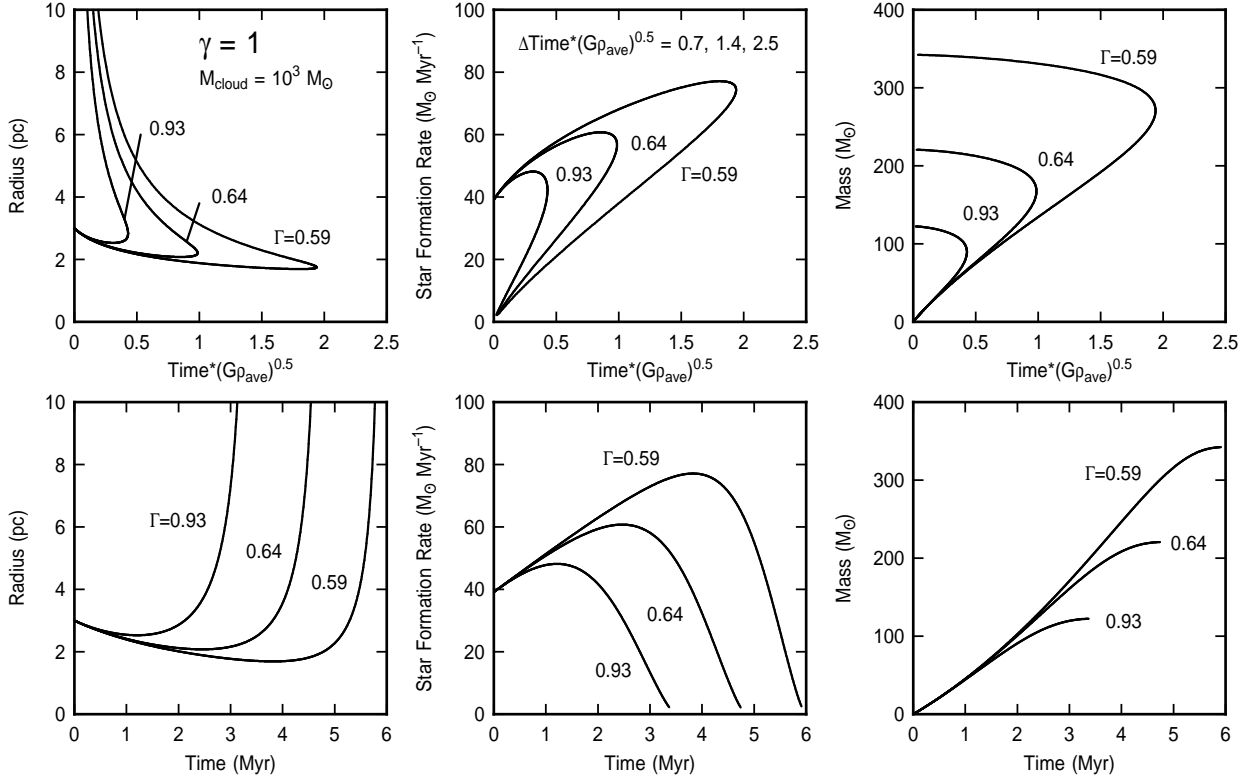}\caption{Models based on the Huff \& Stahler (2006)
formulism with energy input from stars. The bottom panels show cloud
radius, star formation rate, and cluster mass versus the absolute
time, and the top panels show these quantities versus the relative
time, which is the absolute time divided by the instantaneous
dynamical time for the average density inside the cloud. The numbers
in the top middle panel are the product of the instantaneous
crossing rate at the peak in the star formation rate (and at the
peak density) multiplied by the duration of star formation, which is
the time interval between the half-peaks in the star formation rate
shown in the lower middle panel. These models assume the luminosity
of the cluster increases linearly with cluster mass ($\gamma=1$) and
they assume the initial cloud mass is $10^3$ M$_\odot$. This would
be analogous to the Taurus clouds or other regions of low-mass star
formation. The total duration of star formation can be long in
absolute terms, so fairly old stars can be present in such a region
(as indicated by the long absolute times in the lower panels), but
the main activity is always finished in only a few instantaneous
crossing times, as indicated by the small values of $\Delta {\rm
Time}\*\left(G\rho\right)^{0.5}$ in the top middle
panel.}\label{fig:huffr}\end{figure}

\begin{figure}\epsscale{1}
\plotone{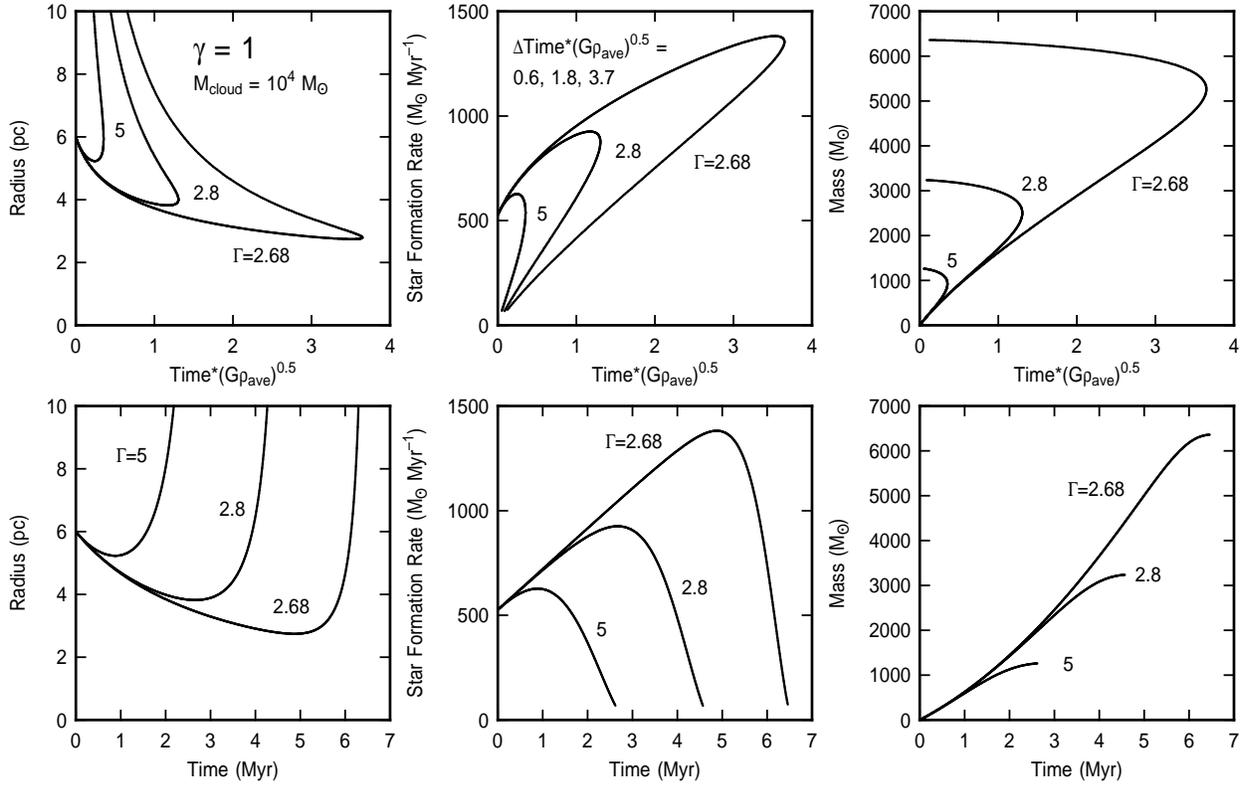}\caption{The same cluster formation model as in Fig.
2 except for a cloud mass of $10^4$ M$_\odot$. The collapse of
massive clouds is difficult to turn around by non-ionizing
radiation, which is the case for this $\gamma=1$ run.  This is
unrealistic, however, because such a massive cloud will form
high-mass stars and then $\gamma>>1$.  $\Gamma=2.68$ is an extreme
case because it forms too many stars (a 64\% final efficiency).
}\label{fig:huffp}\end{figure}

\begin{figure}\epsscale{1}
\plotone{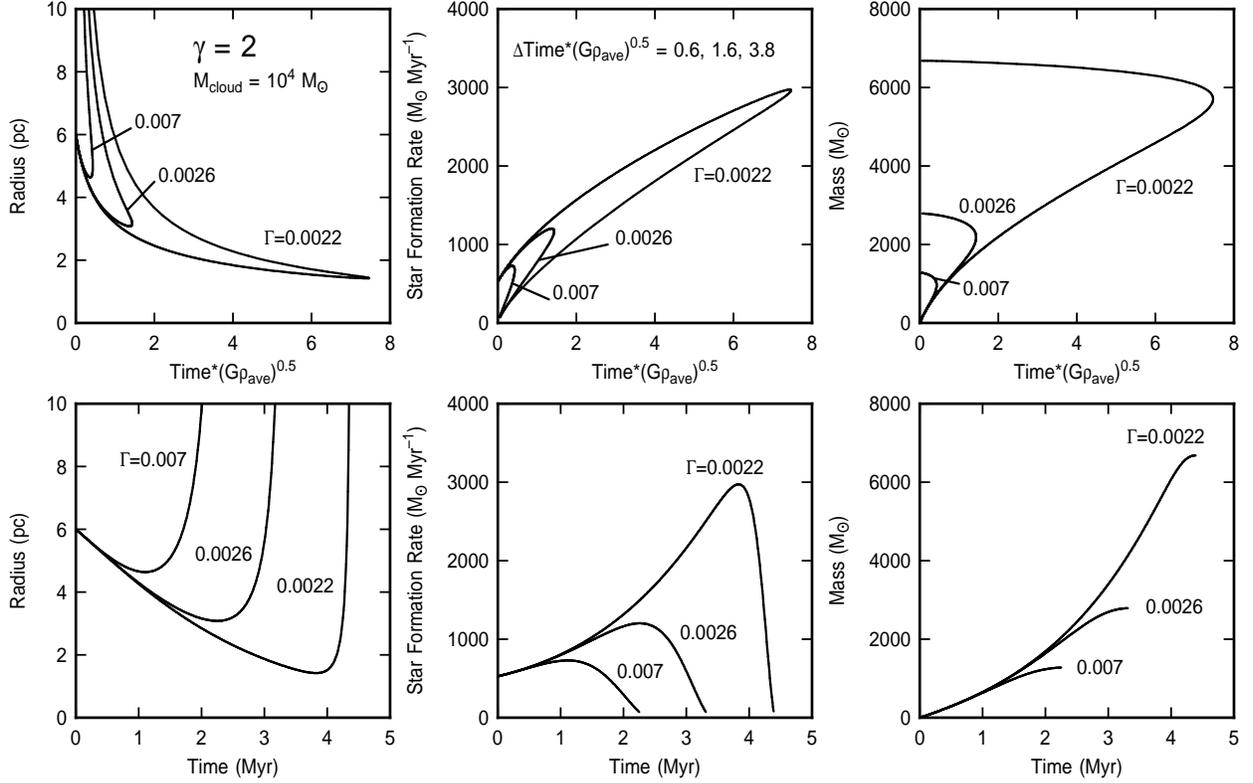}\caption{The cluster formation model with the same
high mass as in Fig. 3, except now with a sensitive relation between
cluster mass and luminosity using $\gamma=2$.  This model applies to
high mass regions that form O-type stars. As in the other models,
the duration of star formation is always short in terms of the
instantaneous crossing time, but the total age span for the stars
can be large in absolute terms. $\Gamma=0.0022$ is an extreme case
because it forms too many stars (a 67\% final
efficiency).}\label{fig:huffq}\end{figure}

Figures \ref{fig:huffr}-\ref{fig:huffq} show the results. In Figures
\ref{fig:huffr} and \ref{fig:huffp}, $\gamma=1$ and the cloud mass
is $10^3$ M$_\odot$ and $10^4$ M$_\odot$, respectively. In Figure
\ref{fig:huffq}, $\gamma=2$ and the cloud mass is $10^4$ M$_\odot$.
The starting radii are $R(t=0)=3$ pc for the first case and 6 pc for
the second two cases; $R_{min}=0.1$ pc. In the bottom panels, the
radius, stellar mass, and star formation rate are plotted versus the
absolute time, and in the top panels these quantities are plotted
versus the relative time, which is the absolute time divided by the
instantaneous dynamical time, $\left(G\rho\right)^{-1/2}$. The
different values of $\Gamma$ are plotted in separate curves, as
indicated.  In each case, the radius decreases at first, as in the
Huff-Stahler solution, because there are few young stars to add
turbulent energy. When the density and star formation rate reach a
sufficiently high value, the energy input rate from stars begins to
exceed the turbulent energy loss rate and the radius increases. Then
star formation slows down because of the decreasing density. Lower
$\Gamma$ cases produce higher stellar masses.

The duration of the most active phase, which is taken to be the full
width at half maximum in the lower middle panel of Figures
\ref{fig:huffr} to \ref{fig:huffq}, is only several instantaneous
crossing times even if the age range in the cluster is many
instantaneous crossing times and a large total value in absolute
time. In Figure \ref{fig:huffr}, for example, the width of the
highest star formation peak is $\sim5.3$ Myr, which is $\sim1.4$
times the age at the peak. In the top middle panel the age at the
peak is 1.8 instantaneous crossing times. Multiplying the relative
time by the fractional total time gives 2.5 crossing times measured
at the peak density for the FWHM duration of star formation. The
relative durations of all three curves are given in the top middle
panel in order of decreasing $\Gamma$. Figures \ref{fig:huffp} and
\ref{fig:huffq} also give the relative durations. For these two
cases, the lowest $\Gamma$ has unrealistically strong star formation
because the final efficiencies exceed 60\%; the middle $\Gamma$ are
best, giving $\sim30$\% efficiency. The corresponding FWHM durations
of star formation are 1.8 and 1.6 crossing times at peak density. In
terms of absolute time, there are old stars present dating back to
when the cluster was young, which can be several million years. The
case closest to a massive dense cluster like 30 Dor is in Figure
\ref{fig:huffq} with $\Gamma=0.0026$, for which a $10^4$ M$_\odot$
cloud produces a $\sim3\times10^3$ M$_\odot$ cluster in a burst
lasting $2.6$ Myr and $1.6$ crossing times at the highest density.

This example is a crude model for the formation and disruption of a
cluster, but it illustrates the point also made in sections
\ref{sect:bigfast} and \ref{sect:smallfast} that star formation can
be fast in terms of the instantaneous dynamical time, even though
the absolute rate varies from slow at the beginning, to fast in the
densest phase, to slow again after the disruption.

A second conclusion to be made from this analysis is that the
conceptual difference between star formation rate and instantaneous
luminosity is important. The luminosity is not proportional to the
star formation rate but to the integral of the star formation rate
over time. So any point in the evolution where the stellar
disruptive luminosity balances the turbulent dissipation rate is
quickly passed as stars continue form.  There is no stable state
because stars keep forming and the luminosity keeps increasing even
when there is a temporary equilibrium. V\'azquez-Semadeni et al.
(2005) and Bonnell \& Bate (2006) also note the lack of stable
equilibria in cluster-forming cores.

Li \& Nakamura (2006) and Nakamura \& Li (2007) took a different
approach. They ran realistic MHD simulations (although without
magnetic diffusion) that generate quasi-stable equilibria through
protostellar wind feedback. The latter paper gets a star formation
rate of a few percent of the cloud mass in each free fall time and
it maintains this rate for 1.5 initial free fall times ($t_{ff}=1.2$
Myr). The equilibrium state is maintained for the last one free fall
time, although it may have been able to continue longer if the code
ran longer. After $1.5t_{ff}$, 80 stars formed. This model is like
the one shown in Figure \ref{fig:huffr} in the sense that it is a
low mass cloud ($\sim10^3$ M$_\odot$) without ionization.  In Figure
\ref{fig:huffr}, the duration of star formation is also about 1.5
instantaneous free fall times, although there is no equilibrium.
This distinction between models raises an important point. In our
figures, the collapse turns around because of intense stellar
pressures at some high density, where the star formation rate is
high. Viewed in a narrow time interval around this turning point,
one dynamical time wide, there is an equilibrium. However, stars
continue to form and the balance of forces continues to build in
favor of cloud dispersal. Soon the cloud expands and star formation
slows down. This all happens in less than a few crossing times. The
same turnaround might happen in the Nakamura \& Li models: as stars
continue to form, their collective winds and radiation should
continue to agitate the gas and eventually overcome the total
dissipation rate, which is somewhat fixed for a constant average
density and Mach number. Their simulated cloud should then expand
and the star formation rate should decrease.  It would seem to take
some tuning to maintain an equilibrium for much longer than 2 to 4
$t_{ff}$ because the collective effects of winds and radiation
should then be quite influential. One way to tune the result would
be to turn off the winds after a short time in each star. A second
point of comparison is that for most of the present paper we are
concerned with massive clusters, and then ionization destroys and
displaces a high fraction of the core gas soon after an O-type star
forms. Nakamura \& Li do not consider this type of energy input, and
even our models in Figure \ref{fig:huffq} do not have a high enough
$\gamma$ to fully account for the sensitivity of energy input to the
presence of massive stars.

Another important effect might be the lack of magnetic diffusion in
the Nakamura \& Li (2007) models (their earlier, 2D, simulations had
magnetic diffusion). Without magnetic diffusion, there is no
possibility of rapid compression-induced triggering, which is an
effect described in their earlier papers and studied again in
Section \ref{sect:comp} below. Star formation triggered by
compression near the outflows could increase the efficiency per free
fall time considerably and lower the overall timescale of the active
phase. It could even remove the impression that there is a
quasi-equilibrium if the phase of gravity/wind force balance becomes
short-lived. Such wind-induced triggerings have apparently been
observed (Barsony 2007). In addition, even in their paper with
magnetic diffusion, Nakamura \& Li (2005) do not consider the
accelerated collapse that might arise from magnetic diffusion rates
proportional to a power of the density greater than 0.5 (Sect.
\ref{sect:magdiff}).

Krumholz, Matzner \& McKee (2006) developed a detailed analytical
model of cluster formation with essentially the same results as
shown here, although their conclusions differed. They considered
spherical self-gravitating clouds with energy input from ionization
by massive stars.  Each generation of star formation in their model
is rapid when measured on a dynamical timescale. Their figures show
oscillations on timescales of 0.5-1 crossing times, and each
oscillation is a generation of stars. Thus their model agrees with
the short timescale for star formation and cloud disruption
discussed in the present paper. However, their model allows the
energized debris from one generation to recollect at the same
position and make more stars later (because everything is
spherically symmetric).  Real GMCs are more filamentary (e.g., Koda
et al. 2006) and star formation at one location cannot easily
influence GMC turbulence or support at a distant point in the same
cloud. Instead, star formation pushes on the gas in its immediate
neighborhood, causing that part of the envelope to move aside, and
at the same time it triggers new star formation in the compressed
region (triggering was not included in their model). Recall that our
application of the Huff \& Stahler (2006) model was only for cloud
cores of modest total mass, not for whole GMCs. Cloud cores are
spherical and somewhat easily disrupted by a single star formation
event, as shown in Section \ref{sect:huff}, while GMCs are elongated
with remote parts that are not so easily disrupted by the same
event.

Krumholz, Matzner, \& McKee also assume that energy from each
generation of star formation affects all of the GMC mass at once,
and through continuous boom and bust oscillations, the total
lifetime of the cloud can be extended. Our view is different. GMCs
show only localized disruption from massive clusters, involving
primarily the core mass ($\sim10^4$ M$_\odot$) and some triggering
in nearby parts of the cloud, with slow and quiescent star formation
elsewhere. Something other than star formation has to support the
remote diffuse parts, and observations suggest this is a combination
of modest turbulence with a strong (sub-critical) magnetic field
(e.g., Cortes, Crutcher, \& Watson 2005). This is the standard model
as far as the low-density envelope is concerned.

Our primary point about rapid star formation is that a whole cloud
begins star formation rapidly somewhere inside of it, and that part
ends star formation rapidly as well. This is unlike the standard
model, which would introduce a delay everywhere of some $10t_{dyn}$
because of slow ambipolar diffusion. For the star-forming part of
the cloud, we agree with the rapid timescale of the Huff \& Stahler
(2006) model. For the rest of the cloud, we agree with the longer
timescale of the Mouschovias (1991) model, provided it is recognized
that the cloud moves around every few crossing times because of
pressure from star-forming cores. Further discussion on long-term
cloud evolution is in Section \ref{sect:dest}.

\subsection{Turbulence Compression and Enhanced Magnetic Diffusion
in Cloud Envelopes}
\label{sect:comp}

Turbulent fragmentation as a model for star formation (see review in
Mac Low \& Klessen 2004) applies best to cloudy regions that are not
collapsing already.  Once collapse begins, the dynamics of the
collapse takes over using the initial conditions from the turbulent
state (e.g., power-law power-spectrum of velocities, with
hierarchical filaments and clumps; Li et al. 2004; Bate \& Bonnell
2005; Jappsen et al. 2005; Tilley \& Pudritz 2005; Martel et al.
2006). Turbulent fragmentation was originally envisioned as a way to
get high densities inside clouds, considering that star formation is
more rapid in the compressed regions than in the cloud as a whole.
Another aspect is also important, and that is the enhanced expulsion
of magnetic flux from the compressed gas (e.g., Nakamura \& Li 2005;
Kudoh \& Basu 2007). The point is that slow ambipolar diffusion at
the low average density of a cloud or GMC envelope is not relevant
for star formation. Diffusion is relatively fast in the clumps where
stars actually form.

Compression enhances diffusion by changing the force density balance
in a magnetically critical cloud from one where $g\rho\sim
B^2/\left(4\pi R\right)\sim\rho n_i\alpha_{in}R\omega_{diff}$ for
gravitational acceleration $g\sim G\rho R$ and cloud radius $R$, to
one where $g_c\rho_c<B_c^2/\left(4\pi L\right)\sim\rho_c
n_{i,c}\alpha_{in}L\omega_{diff,c}$ in the clumps. Subscript {\it c}
represents the compressed clump state, $L$ is the compressed clump
size, and $\omega_{diff}$ is the magnetic diffusion rate.
Self-gravity is written here as relatively unimportant during the
initial compression, although this would not always be the case.
Writing the clump ionization fraction as $x_c =n_{i,c}m/\rho_c$ for
mean molecular weight $m$, we get a magnetic diffusion rate
\begin{equation}
\omega_{diff,c}={{B_c^2m} \over {\left(\rho_c
L\right)^2x_c\alpha_{in}}}.\end{equation} The column density does
not change much with fast lateral compression ($\rho R\sim\rho_c L$)
but the field strength does, by flux freezing, as $B_c\sim B
\left(R/L\right)$, thus
\begin{equation}
{{\omega_{diff,c}}\over{\omega_{diff}}}\sim
{{R^2x}\over{L^2x_c}}>>1.
\end{equation}

This enhancement factor for the diffusion rate is larger than the
time factor during which the turbulence-compressed state is
maintained, so there is a net flux loss from the clump. The duration
of the pre-compressed state is $\tau=R/v_A$ for Alfv\'en speed
$v_A=B/\left(4\pi\rho\right)^{1/2}$ initially comparable to the
virial speed $\left(gR\right)^{1/2}$. The duration of the compressed
state is $\tau_c=L/v_{A,c}$ where $v_{A,c}\sim v_A
\left(R/L\right)^{1/2}$. Thus the ratio of durations is
$\tau_c/\tau=\left(L/R\right)^{3/2}$. Multiplying this by the ratio
of diffusion rates, we get the relative enhancement of flux loss as
\begin{equation}
{{\omega_{diff,c}\tau_c}\over{\omega_{diff}\tau}}\sim
{{R^{1/2}x}\over{L^{1/2}x_c}}\sim\left({R\over
L}\right)^{0.5+\kappa}>>1
\end{equation}
for ionization fraction varying with density as $x\sim n^{-\kappa}$
and $n$ inversely proportional to size during the compression. In
simulations by Nakamura \& Li (2005), $\kappa\sim0.5$ and they find
enhanced flux loss. The compression-induced flux loss is larger if
$\kappa\sim1$ in the dense state, as suggested in the Section
\ref{sect:magdiff}.

For a magnetically critical cloud, $\tau\sim t_{dyn}$ and
$\omega_{diff}\tau\sim t_{dyn}/t_{diff}\sim1/10$. If
$\left(R/L\right)^{0.5+\kappa}>10$, which is reasonable, then
$\omega_{diff,c}\tau_c>1$ and the clump diffusion time is less than
the duration of the compressed state. This means that a high
fraction of the magnetic flux will diffuse out. In this case,
turbulence compression not only makes the dense regions but it also
forces so much magnetic flux from them that they can become
supercritical in a crossing time and collapse quickly into stars.
This triggering process is much faster than the same gas would have
evolved on its own from an initially sub-critical state.

The regions where turbulence-enhanced compression and diffusion
should be important include marginally stable GMCs envelopes and
non-collapsing central regions, in addition to diffuse and
low-pressure molecular regions that are sub-critical on average.
Compression-enhanced diffusion should also be important in GMC
envelopes where HII regions and other pressures trigger star
formation. The compression has to be strong enough ($R>>L$) to make
$\omega_{diff,c}\tau_c>1$. We consider in Section \ref{sect:imf}
whether the IMF should be different in turbulence-compressed regions
than in collapsing supercritical cores, and possibly different again
for star formation that follows the standard model of slow ambipolar
diffusion before collapse.

\subsection{Enhanced Magnetic Diffusion in Cloud Cores}
\label{sect:magdiff}

Microscopic changes should also play an important role in the rapid
collapse of GMC cores, and they should aid with accelerated
diffusion in the turbulence-compressed clumps of GMC envelopes. An
essential consideration is how rapidly the ratio of the magnetic
diffusion time to the dynamical time decreases at higher density.
For typical cosmic ray ionization rates, the density scaling for the
electron fraction changes from $x\propto n^{-1/2}$ to $n^{-1}$ when
charge exchange replaces dissociative recombination for the
neutralization of ionic molecules, and electron recombination on
neutral grains replaces dissociative recombination with ionic
molecules (Elmegreen 1979; Draine \& Sutin 1987). For cosmic ray
ionization rates typical of the solar neighborhood, this change
occurs at $n\sim10^5$ cm$^{-3}$. The density scaling is important
because the ratio of the diffusion time to the dynamical time drops
faster for steeper scaling laws. For example, Basu \& Mouschovias
(1995) showed that dynamical evolution is faster when $x\sim
n^{-2/3}$ than when $x\propto n^{-1/2}$. Hujeirat, Camenzind, \&
Yorke (2000) considered various density dependencies and found that
if the power exceeds $2/3$, the time for an initially subcritical
core to start collapsing dynamically is equal to the initial free
fall time. The $n^{-1/2}$ to $n^{-1}$ scaling transition at
$n\sim10^5$ cm$^{-3}$ is shown in Figure 1 of Elmegreen (1979) and
in Figures 1-6 of Umebayashi \& Nakano (1990).

Also at high density, the waves generated by cosmic ray streaming
instabilities damp faster than their growth rate and cosmic rays
stream freely along the field lines. The cosmic ray density drops
sharply at this point. This drop is shown in Figure 1 of Padoan \&
Scalo (2005), where for dark cores it also occurs at a density of
$\sim10^5$ cm$^{-3}$. A sudden drop in the cosmic ray ionization
rate inside the dense parts of clouds would lead to an even greater
drop in the ionization fraction.

Further loss of magnetic support at this density should arise
because of changes in the grain population. Charged grains
contribute substantially to the magnetic support of neutral
molecules, and small grains dominate the viscous cross section.
However, observations suggest that PAH molecules and small grains
disappear in very dense clouds (Boulanger, et al. 1990). Depletion
could cause small grains to grow.  Omont (1986) suggests the
depletion time onto grains is $10^{10}/n$ years, which is smaller
than the dynamical time, $\left(G\rho\right)^{-1/2}$, when
$n>3\times10^4$ cm$^{-3}$. Depletion also removes ionic metals which
lowers the ionization fraction.  Also at about this density, grain
coagulation reduces the number of charged grains and this too
reduces grain coupling to neutrals (Flower, Pineau des For\^{e}ts,
\& Walmsley 2005). Further coupling loss arises because large grains
lose their field line attachment (Kamaya \& Nishi 2000). All of
these microscopic effects speed up star formation at $n\sim10^5$
cm$^{-3}$ for Solar neighborhood conditions by allowing the magnetic
field to leave the neutral gas more quickly. Only a few of these
effects have been included in MHD simulations.

\subsection{Slow Protostellar Motions in Rapidly Evolving Clouds}
\label{sect:motions}

Newborn protostars that form in magnetic turbulent gas should move
slower than the virial speed for two reasons. First, the magnetic
field provides some support to the cloud, so most of the gas moves
at sub-virial speeds anyway. Second, protostars that form in
turbulence-shocked regions will have the average speed of the two
colliding streams; the component of the velocity perpendicular to
the shock will cancel. If magnetic energy, turbulence and
self-gravity have comparable energy densities, then the turbulent
speed is $\left(1/2 \right)^{1/2}$ of the virial speed. Colliding
flows reduce the final protostar speed by another factor of
$\left(2/3\right)^{1/2}$, on average, so the net reduction is a
factor of $\left(1/3\right)^{1/2}=0.58$. Thus protostars should
appear to be moving relatively slowly. Observations by Belloche,
Andr\'e, \& Motte (2001), Di Francesco, Andr\'e \& Myers (2004),
Walsh, Myers \& Burton (2004), J{\o}rgensen et al. (2007) and Walsh
et al. (2007) show slow motions for pre-stellar cores.

The slow birth motion of pre-stellar cores implies that the
protostars they eventually make will be subvirial and sink to the
center of the cloud, increasing the star-to-gas mass fraction there
and decreasing the required total efficiency for cluster
self-binding (Elmegreen \& Clemens 1985; Pinto 1987). Patel \&
Pudritz (1994) proposed that the cold stellar component in an
embedded cluster would collapse inside the gaseous component by a
two-fluid instability. If the initial protostar speed is $v_i$ and
the virial speed is $v_v$, then the formation efficiency that
produces a 50\% stellar mass fraction after protostar settling is
$0.5\exp\left(-0.75\left[1-v_i^2/v_v^2\right]\right)$ for an
isothermal cloud (Verschueren 1990).  The formation efficiency for
cluster binding with instantaneous gas removal is therefore 50\% for
$v_i=v_v$ and 30\% for $v_i=0.58v_v$. It is smaller for slow gas
removal (Lada, Margulis, Dearborn 1984), and smaller still if some
stars escape leaving a tighter cluster in the core (Boily \& Kroupa
2003).

Before a pre-stellar clump detaches from the magnetic field on which
it formed, its motion will be influenced by the magnetic field. If
the magnetic field in a protostellar clump is critical, or if the
clump forms with a constant mass-to-flux ratio in a cloud where the
average magnetic field is critical, then the field strength in the
clump satisfies, $B_{clump}\sim2\pi G^{1/2} \Sigma_{clump}$ for
clump mass column density $\Sigma_{clump}$ (Sect. \ref{sect:sc}).
The magnetic force per unit volume acting on the clump by the field
lines it drags behind is approximately $F_B=B_{clump}^2/\left(8\pi
R_{clump}\right)\sim G\Sigma_{clump}^2/R_{clump}$. The force per
unit volume acting on the clump by the gravity from the rest of the
cloud is $F_G\sim G\Sigma_{cloud}\rho_{clump}\sim
G\Sigma_{cloud}\Sigma_{clump}/R_{clump}$. Thus the ratio of the
magnetic to the gravitational forces acting on the clump from the
surrounding cloud is
\begin{equation} F_B/F_G \sim \Sigma_{clump}/\Sigma_{cloud}
>>1.\end{equation}
This latter inequality is usually satisfied because protostellar
clumps have low angular filling factors, which means their column
densities are higher than the average cloud column densities around
them. As a result, clumps do not free fall in a cloud until either
their magnetic field lines become detached or their fields diffuse
out. This is one of the reasons why clump motions can be slow.

Magnetic fields should also limit clump accretion from remote parts
of the cloud. The magnetic force per unit volume exerted on the
ambient gas in a cloud is $\sim B_{cloud}^2/\left(8\pi
R_{cloud}\right)\sim G\Sigma_{cloud}^2 / R_{cloud}$. The
gravitational accretion force per unit volume that the clump exerts
on this ambient gas is $\sim GM_{clump}\rho_{cloud}/R_{cloud}^2$.
The magnetic to gravitational force ratio for accreted ambient cloud
gas is
\begin{equation} F_B/F_G \sim
M_{cloud}/M_{clump}>>1.\end{equation} Thus the ambient cloud gas
cannot freely fall onto a clump whose mass is significantly less
than the mass of the whole cloud.

Pre-stellar clump motions and gas accretion onto clumps from remote
parts of the cloud should be restrained by the cloud's magnetic
field if it is close to the critical value. Pre-stellar clumps are
therefore born with relatively slow speeds, and they should keep
these speeds until their field lines detach. The protostars they
form should accrete only from their immediate clump reservoirs or
from closely interacting clumps. These protostars should also move
slowly for a long time, even if they become detached from the field
lines, because the protostars are bound to their clumps by clump
gravity with a stronger acceleration ($G\Sigma_{clump}$) than they
are attracted to the whole cloud ($G\Sigma_{cloud}$), considering
that $\Sigma_{clump}>\Sigma_{cloud}$.  Protostars begin to move
freely only when they become detached from the cloud's field lines
and also destroy the clump that formed them. Before this detachment,
protostars should appear offset from their clump centers with an
equilibrium position that balances clump and cloud forces from
magnetic fields, ram pressure, and gravity. The observation of slow
protostellar and prestellar motions may eventually be used to
determine the magnetic field strength. Faster motions compared to
virial imply weaker fields compared to critical.

\section{The Morphology of Destruction: Triggered Star Formation and
Longevity in Molecular Cloud Envelopes} \label{sect:dest}

Jets, winds, heating, and ionization in dense cluster-forming cores
can compress the existing clumps and produce tiny shells, both of
which may trigger more star formation (Norman \& Silk 1980; Quillen
et al. 2005; Barsony 2007).  If only low mass stars form, the energy
input may not be disruptive and the core might survive for several
crossing times. If high mass stars form, then the core should be
rapidly dispersed. Gas exhaustion also halts star formation. In a
region that forms a bound cluster, nearly half of the gas is
converted into stars and little remains in a dense state for more
star formation.

These two endings for core activity are readily observed. High mass
cores that form O-type stars make compact HII regions in the midst
of dense clusters of lower mass stars. These HII regions clear out
small cavities at first and change the mode of star formation from
one of collapse and turbulence compression to one of triggering at
the cavity edges and in the debris. Low mass cores with no O-type
stars should contain smaller, less energetic bubbles when they are
young (Quillen et al. 2005) and a gradual lessening of extinction
over time as the gas gets used up, rather than an explosive
clearing. The efficiency may reach $\sim30$\% by the clearing time
in both cases (Lada \& Lada 2003). The age of a newly cleared
cluster is typically short, only a few crossing times.

The general speed up of star formation with density implies that GMC
cores are finished before the diffusion time in the envelope. This
is particular true if the GMC envelope is subcritical, which seems
likely (Sect. \ref{sect:sc}). In addition, GMCs in the main disk of
the Milky Way have an average column density equivalent to $\sim10$
mag of visible extinction (Solomon et al. 1987). Because it takes
$\sim4$ magnitudes for a clumpy cloud to significantly shield the
background uv light (McKee 1989; Ciolek \& Mouschovias 1995; Myers
\& Khersonsky 1995; Padoan et al. 2004), there should be
considerable ionization in the envelopes of GMCs. This means the
magnetic diffusion time can be long, many tens of dynamical times.
Thus we have an exception to the highly dynamical picture presented
in the preceding sections: GMC envelopes can be relatively
long-lived.

Envelope longevity appears necessary also from the Zuckerman \&
Evans (1974) constraint, which suggests that CO clouds cannot be
collapsing as a whole, and from the Solomon, Sanders \& Scoville
(1979) constraint, which notes that the inner Galaxy is highly
molecular (Sect \ref{sect:bigfast}).  Dynamical evolution of GMCs
means primarily that they progress toward star formation relatively
quickly and then scatter their envelopes relatively quickly. But it
does not mean that the scattered envelope disappears.

\begin{figure}\epsscale{1}
\plotone{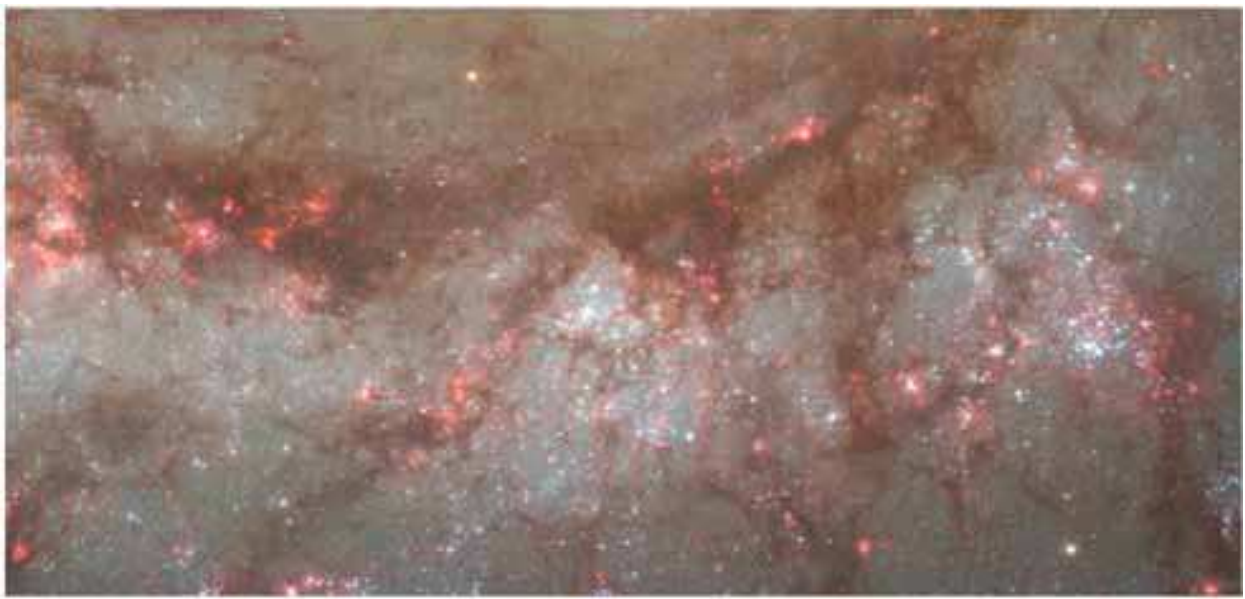} \caption{A section of the southern inner arm
of M51 from the full-resolution Hubble Heritage image. The overall
dimensions are 3.49 by 1.65 kpc. There are 2 giant cloud complexes
in the main dust lane ($10^7$ M$_\odot$). Each has embedded HII
regions, showing that star formation begins very soon after cloud
formation with no significant downstream displacement. The
post-shock flow is mostly from left to right in this figure as the
gas streams along the spiral arm. The feathers of dust clouds below
the main dust lane are twisted remnants of former cloud complexes in
which star formation has dispersed the cores. Many filaments or
ribbons of dust surround complexes of bright blue stars, suggesting
pressurized dispersal. Star formation lingers in the filaments and
in other debris because of triggering from these pressures and
because of parallel collapse along the filaments into dense knots.
The lowest density regions do not show star formation. These low
density regions are presumably the envelopes and shredded debris of
former GMCs. They appear to last a relatively long time; some even
get to the next spiral arm (not shown) [Image degraded for astroph].
}\label{fig:m51blowup}\end{figure}

An example of rapid star formation with slow GMC envelope evolution
is shown in Figure \ref{fig:m51blowup}, which reproduces the
southern part of the inner spiral arm in M51 from the HST Heritage
mosaic. There is a clear progression of star formation morphology
from left (east) to right that matches the expected time evolution
as the gas flows away from the spiral shock. In the east, there is a
large concentrated dark cloud that is part of the dust lane itself.
It measures $1.0\times0.23$ kpc$^{2}$ (assuming the distance is
$9.5$ Mpc from Zimmer et al. 2004) and with an estimated average
visual extinction of 2 mag, contains $10^7$ M$_\odot$. Star
formation occurs throughout this cloud in several places, so there
is no perceptible time delay between cloud formation and star
formation. The giant cloud itself is a ``giant molecular
association'' or ``supercloud,'' and the relatively small
concentrations in it, barely visible at $\sim100$ pc in length,
would be the GMCs. This size and hierarchical morphology is common
in the spiral arms of our Galaxy too (Grabelski et al. 1987). In the
middle of the image there is relatively little in the center of the
dust lane but there is one kpc-long clumpy cloud extending south and
there are several small cloud filaments to the west of it, along
with many small HII regions in the dense knots. The small filaments
make irregular shells, and there are many bright blue stars inside
these shells that could have pressurized them. Further to the west
there is another supercloud with embedded HII regions inside the
dustlane and there are two other kpc-scale filaments south of the
dustlane and aligned perpendicular to it (``feathers'' -- Shetty \&
Ostriker 2006; La Vigne, Vogel \& Ostriker 2006). These filaments
have low-level star formation along their edges. There are also more
blue stellar associations between the filaments, giving the overall
appearance of shells again.  By shells, we do not mean
three-dimensional objects; the in-plane dimensions are much larger
than the gas scale height, so these are more like ribbons or loops
in the plane. The observed progression from left to right in the
figure is the flow direction for gas in the density wave crest. Each
feather has swung out counter-clockwise from the dust lane because
of reverse shear on the inner side of the arm (Balbus 1988; Kim \&
Ostriker 2002, 2006). The time scale for this is several tens of Myr
at a relative speed of $\sim50$ km s$^{-1}$ (according to Kim \&
Ostriker 2006, the spur patterns move at the rotation speed of the
disk relative to the arm).

The figure shows many aspects of the present discussion: the rapid
appearance of cores and stars in giant clouds that form in the dust
lane, the shredding of these clouds downstream, the appearance of
500-pc scale star complexes and their 80-pc knots, which are OB
associations in the classic definition, the diffuse, filamentary and
shell-like nature of the cloudy debris, and the lingering star
formation in the cores of this debris. Most of the dense cloud cores
contain some level of star formation inside or immediately adjacent
to them, and much of the diffuse filamentary gas has little star
formation. Essentially all of the dark clouds should be molecular.
The classical notion that there is a delay between the spiral shock
and star formation is not evident: star formation is immediate in
the superclouds. The offset between the main dustlanes and the blue
light that has long been interpreted as a time delay for star
formation is in fact from cloud distruction following dustlane
emergence and from triggered and lingering second-generation star
formation in the shear-twisted debris.

The time scale for evolution of the filaments, which are apparently
the scattered envelopes of superclouds and GMCs, can be assessed
from this figure. The smallest filaments in the image are $\sim20$
pc wide. For one magnitude of extinction and a depth equal to their
width, they would have an H density of $\sim30$ cm$^{-3}$. The
bigger filaments would have slightly lower average densities because
their extinctions look about the same. These structures resemble
local diffuse clouds although they are probably molecular in M51,
which is molecule-rich. The dynamical time at this density is
$\sim15$ My, which is a good fraction of the spiral arm flow time
represented in the figure. Since most of the filaments contain star
formation inside or adjacent to their cores, the onset of star
formation in the debris appears to operate relatively quickly, on
approximately the dynamical time. This would seem to be impossible
if the clouds are diffuse.  Magnetic diffusion should be slow in the
low-density parts of this gas because they are highly exposed to
ambient starlight, and if the debris came from the disruption of
sub-critical GMC envelopes, it should be sub-critical in filamentary
form too.

The morphology of clouds and HII regions in Figure
\ref{fig:m51blowup} gives a clue to the continued activity on
relatively short time scales. There seem to be two mechanisms for
second generation star formation: direct triggering from HII regions
and other pressures associated with the existing blue stars, and
gravity-driven streaming of gas along the filaments to make dense
cores (e.g., Nakamura, Hanawa, \& Nakano 1993; Tomisaka 1995; Fiege
\& Pudritz 2000). The first of these short-circuits the long
diffusion time by compressing the gas (Sect. \ref{sect:comp}. The
second overcomes magnetic resistance directly by increasing the
mass-to-flux ratio, presumably to the supercritical point. The cores
are well separated and the mass-to-flux ratio increases by the ratio
of filament length divided by filament width, which is a factor of
$\sim10$. Gas that is sub-critical by a factor of 10 in low-density
filaments can become supercritical when it collects to a core, and
then it can collapse even if the rest of the filament has a long
diffusion time. Filament streaming takes a time
$\sim(L/W)\left(G\rho\right)^{-1/2}$ for collection length $L$ on
each side of the core, filament width $W$, and filament density
$\rho$. This time is longer than $\left(G\rho\right)^{-1/2}$ alone,
but the process is still dynamical and it requires minimal magnetic
diffusion before star formation begins in the core.

The low-density sub-critical debris left behind in the filaments
should continue to have a lifetime significantly longer than the
dynamical time because it is sub-critical and highly exposed. It
satisfies the Zuckerman \& Evans constraint and the Solomon, Sanders
\& Scoville constraint. But these constraints have little bearing on
star formation time scales when they are satisfied by diffuse GMC
debris.  The fate of this gas depends more on supernovae and the
clouds' pending impact with the next spiral arm.

We conclude that most or all of the gas that is strongly
self-gravitating evolves toward star formation on a dynamical time,
whether it is forming a first generation of stars in a supercloud or
subsequent generations in the debris. A high fraction of the volume
of the ISM is in the form of diffuse gas, which may evolve in
isolation more slowly than the internal dynamical time considering
the likely subcritical fields and strong ionic attachment to these
fields. However, this diffuse gas is forced to evolve in other ways
by supernovae and stellar pressures in its vicinity. Sometimes these
pressures trigger star formation, speeding up magnetic diffusion and
collapse locally, and sometimes they move the gas into tenuous
shells and filaments that eventual get ionized and disperse. The
full image of M51 indicates that many of the diffuse extinction
clouds make it all the way to the next arm, so they last a
relatively long time if they are left alone.

In a recent paper, Mouschovias, Tassis, \& Kunz (2006) reviewed the
observations which historically suggested slow ambipolar diffusion
time scales for GMC evolution. The present model agrees with their
assessment that cloud envelopes are subcritical and slow to evolve,
and that cloud cores are supercritical and collapsing. However, they
believe the cores form slowly on the ambipolar diffusion time and we
suggest they form quickly because they are close to critical from
birth or they are compressed. They also referred to the separation
between dustlanes and HII regions in spiral density waves as
evidence for long time scales prior to star formation, but we have
shown for M51 that there is no such delay.  The difference between
these two views lies entirely in the different initial conditions
for core formation, not in the theory of ambipolar diffusion and
collapse. We also differ in our consideration of cloud disruption
and secondary star formation, which can be rapid because of high
stellar pressures.

There is no single mode of star formation but several, starting with
what might be called a primary mode that begins with large scale
gravitational instabilities in spiral arm dustlanes and elsewhere in
the ambient ISM, continuing with triggered star formation during the
disruption of the cores and envelopes in these primary clouds, and
lingering further still with the dynamical collapse of filamentary
debris and more pressurized triggering during envelope dispersal. We
believe considerable evidence supports a picture where the onset of
star formation in almost all clouds is at the dynamical rate, not
the slower ambipolar diffusion rate. The evidence also suggests that
the complete destruction of clouds can be considerably slower,
giving the molecules long total lifetimes. Thus cloud evolution
consists of a mixture of rapid and slow processes. These are nicely
mapped out as a time sequence in the downstream flow from a strong
spiral arm. In galaxies with weak or no stellar spirals, the same
phases of cloud evolution should occur, but they will be mixed
together in space as there is no global trigger for the first stage.

\section{IMF Variations for the 3 Modes of Star Formation}
\label{sect:imf}

The previous sections presented evidence that star formation has
three distinct modes: (1) rapid collapse for small stellar groups
and single stars in turbulence-compressed regions, which may mix
into clusters or remain dispersed, (2) rapid collapse of
supercritical cores that are born with near-critical field strengths
as a result of larger-scale galactic processes, and (3)
supercritical collapse of single stars and clusters following slow,
diffusion-limited contraction in an initially subcritical cloud. The
latter is the standard model but appears to take too long for the
general case and to give the wrong proportion of pre-stellar and
stellar cores. The disruption of star-forming clouds was shown to
involve another type of star formation, which is a variant of the
first mode: triggered star formation in cloud envelopes and in
debris from previous generations of stars. Examples of these modes
were given and the whole evolutionary scenario was illustrated using
HST Heritage images of M51.

These modes differ in fundamental ways so it is natural to expect
some differences in the properties of stars they produce. The IMF,
for example, could differ between quiescent regions in mode (3) and
large-scale collapsing regions in mode (2). Turbulent fragmentation
in mode (1) would seem to give a different IMF also. Binary
fractions, mass segregation, efficiencies and other properties of
star formation might differ as well. Remarkably, simulations with
extremely diverse conditions, ranging from pure collapse with no
magnetic fields (Bonnell \& Bate 2006) to highly constrained and
localized with strong fields (Tilley \& Pudritz 2005) all give about
the same IMF slope at intermediate mass. Simulations have not yet
sampled out to high mass. The turnover at low mass depends on the
assumptions of the model, and although the models can be tuned to
give the right result, the origin of the turnover is not understood
yet (compare, for example, the different reasons for a turnover in
Padoan \& Nordlund 2002, Martel et al. 2006, and Jappsen et al. 2005
or Bonnell, Clarke \& Bate 2006).

Padoan \& Nordlund (2002) and Padoan et al. (2007) discuss how
magnetic turbulent compression with Kolmogorov-type scaling laws
between velocity and length can partition the gas into pieces that
have the Salpeter mass function at intermediate mass. Di Fazio
(1986) and Elmegreen (1993, 1997) suggested a slightly different
scenario where the IMF slope comes not only from instantaneous mass
partitioning but also from differential collapse rates, which
steepen the slope for the time-integrated population. Bonnell,
Larson, \& Zinnecker (2007) present a case for collapse without
magnetic restraints, where gas can move freely relative to the dense
cores and protostellar masses grow by competition accretion. We
discussed how critical magnetic fields should limit this scenario in
Section \ref{sect:motions}.

Turbulent fragmentation theories would seem to apply best to the
first star formation mode. Competitive accretion seems to apply best
to the second mode, i.e. to supercritical, collapsing,
cluster-forming cores or parts of cores, where the magnetic field is
relatively weak, clump motions are relatively unconstrained, and
collapse motions dominate broad-spectrum turbulence for the
dynamics. If this is the case, then we can assess what the possible
IMF differences might be.

Bonnell, Larson, \& Zinnecker (2007) show that in freely collapsing
models, stellar masses grow mostly by accretion, sometimes from far
away, and that stellar interactions and sub-cluster ejection limits
the accretion for what turn out to be the low mass stars (e.g. Bate
\& Bonnell 2005). Bonnell \& Bate (2006) note that competitive
accretion works well for the high mass stars. The high masses of the
highest mass stars can even run away in this model, because the
accretion rate increases with mass (e.g., Martel et al. 2006). Also
in dense cluster-forming cores, pre-stellar clumps might coalesce to
make more massive stars (Peretto, Andr\'e, \& Belloche 2006). Thus
mode 2 would seem to be able to produce an excess of massive stars
if the density gets high and the gas reservoir is large, as in a
massive cluster. This would make the mode 2 part of the IMF somewhat
shallow. The mode 1 part of the same cloud would presumably not have
such coalescence and runaway properties: each star is forced to
accrete from its immediate neighborhood because of magnetic
stresses. Then the IMF might be slightly steeper. The composite IMF
in a cluster that contains both a supercritical collapsing inner
core and a turbulence-compressed outer core or envelope would have
an IMF gradient and an intermediate average IMF. The outer IMF would
be steeper than the inner IMF, and the average would depend on what
fraction of the total mass was in the supercritical collapsing
state.  Such a variation would naturally account for mass
segregation at stellar birth and for the slight trend toward
shallower IMFs with increasing density and pressure (Elmegreen
2004).

The Padoan \& Nordlund (2002) model assumes the clump mass that
forms in a compressed layer is proportional to $\rho L^3$ for
compressed density $\rho$ and layer thickness $L$.  The uncompressed
values are $\rho_0$ and $L_0$. By mass and flux conservation, $\rho
L=\rho_0 L_0$ and $B\rho=B_0\rho_0$, so the magnetic field dominates
the layer pressure, giving $\rho v=\rho_0 v_0$ for velocity
dispersions $v$ and $v_0$.  With space partitioned as $P(k)d\log
k=k^3d\log k$ and velocity correlated in the surrounding cloud as
$v_0/v_c=\left(L_0/R\right)^\alpha$ for cloud radius $R$, they get a
power law mass function $f(M)d\log M=P(k)d\log k$ for $M=\rho
L^3=\rho_0L_0L^2
\propto\rho_0L_0^3\left(v/v_c\right)^2\left(R/L_0\right)^{2\alpha}$.
If the initial scale for the compression, $L_0$, is identified with
the inverse wavenumber of the turbulence, $k^{-1}$, then $M\propto
k^{2\alpha-3}$. Combining this with $P(k)$, the result is
$f(M)\propto M^{3/\left(2\alpha-3\right)}\propto M^{-1.33}$ for
$\alpha=0.37$, from the velocity power spectrum. The analogous
result for filaments is slightly different. Assuming again that
$M=\rho L^3$ for filament width $L$, we now have conservation laws
$\rho L^2=\rho_0L_0^2$ and $BL^2=B_0L_0^2$, which still gives
$\rho\propto B\propto v$ by pressure balance, but then the mass
becomes $M=\rho_0L_0^2L=\rho_0L_0^3\left(v/v_c\right)^{1/2}
\left(R/L_0\right)^{\alpha/2} \propto k^{\alpha/2-3}$. With the same
space partitioning, $P(k)\propto k^3$, this gives an IMF for
filaments of $f(M)\propto M^{9/\left(\alpha-6\right)}\sim
M^{-1.60}$, which is slightly steeper than the Salpeter function.

If the compressed layers produce a number of star formation sites
proportional to the area layer, namely $\left(L_0/L\right)^2$, and
the cylinders produce a number of sites proportional to the length,
$L_0/L$, then both of these IMFs becomes $f(M)\propto M^{-1}$. This
can be seen by using $P(k)=k^3 \left(L_0/L\right)^2d\log k$ for the
layer and $k^3 \left(L_0/L\right)d\log k$ for the filament, i.e.,
counting not just compression space in the original cloud (the $k^3$
term) but also the number of sites per compressed region. In the
first case, $k^3\left(L_0/L\right)^2=
k^3\left(v_c/v\right)^2\left(R/L_0\right)^{2\alpha}\propto
k^{3-2\alpha}$ and in the second case $k^3\left(L_0/L\right)=k^3
\left(v_c/v\right)^{1/2}\left(R/L_0\right)^{\alpha/2}\propto
k^{3-\alpha/2}$.  Because $M\propto k^{2\alpha-3}$ and
$k^{\alpha/2-3}$ in the layer and filament geometries, the IMFs are
simply $M^{-1}$. This is the usual result for hierarchical
structure.

If the turbulence-compressed layers and filaments fragment into star
formation sites in this way, giving $f(M)\propto M^{-1}$ in both
cases, then the instantaneous mass spectrum from turbulence
fragmentation would be shallower than the observed IMF.  Also in
this case, we should observe linear strings of the youngest
pre-stellar condensations and protostars along gaseous filaments
with a number of condensations increasing with filament length.  We
should see two dimensional arrays of condensations and protostars
inside compressed layers with the number of condensations increasing
with the area of the layer.  To get the Salpeter IMF or something
steeper requires additional physics. Elmegreen (1993, 1997)
suggested this was the mass dependence of the dynamical timescale
for evolution of gas into protostars: in a hierarchically structured
cloud, the smaller pieces, which have lower masses, tend to be
denser on average, and to have shorter dynamical times. Thus an
instantaneous mass function from turbulence steepens into the IMF as
proportionally more low mass clumps and stars form by turbulence
compression and other dynamical processes.

The IMF is likely to be much more complicated than either theory
predicts, and possibly the result of a combination of effects.
V\'azquez-Semadeni, Kim \& Ballesteros-Paredes (2005) for example,
found that the collapsing mass increases as the relative magnetic
field strength increases. This implies that the IMF in sub-critical
regions of clouds, such as GMC envelopes and diffuse regions, might
be shallower than the IMF in critical or super-critical regions
where the field strength is relatively low. Such a trend would
counter the mass segregation gradient discussed above.

\section{Summary}
\label{sect:sum}

The previous sections considered many facets of star formation that
all fall under one basic model consisting of three tenets: (1)
clouds of various origins are hierarchically structured as a result
of turbulence and self-gravity, (2) their densest parts evolve
toward star formation at about the local dynamical rate, and (3)
their low-density envelopes disperse as a result of this star
formation and survive in pieces for several dynamical times,
possibly forming stars in multiple generations.

Cloud formation in the first step includes compression and
gravitational collapse in spiral shocks, swing-amplified spiral
arms, expanding shells and other dynamical structures that form by
turbulence and stellar pressures on a wide range of scales. The
formation and initial evolution occurs on the dynamical time scale,
and in absolute terms, this can be large, several tens of millions
of years, or small, a million years or less, depending on the
pre-cloud density. The low density parts of these clouds, which are
most directly exposed to background starlight, are most strongly
tied to the magnetic field and evolve more slowly in relative terms
than the dense, optically-thick cores.  Core evolution starts close
and stays close to the magnetically critical state, and with
relatively little delay, becomes magnetically supercritical. At this
point star formation proceeds at a high rate, which is still close
to the dynamical rate but now at a high density. Star formation
appear to accelerate as the cloud density increases, but observers
at each stage would see it operating at only a few times the current
dynamical rate even though relatively old stars are present. Star
formation disrupts the core relatively quickly, with no intermediate
stage of dynamical equilibrium.

The envelope becomes disrupted too, following core star formation,
but if there is a relatively large internal magnetic field and a
relatively slow diffusion rate, then it can be pushed to the side,
broken, and dispersed without collapsing into stars immediately. The
envelope also has star formation during this whole process, and it
can be supercritical in small regions where turbulence compression
and external pressures accelerate the diffusion rate. Star formation
finally stops when all of the residual gas is converted back into a
low-density, weakly self-gravitating state by starlight heating,
ionization, and evaporation.

Cloudy debris that is relatively isolated and in a low pressure
environment, such as the Taurus clouds, can form stars by the
standard model, one at a time by quasi-equilibrium magnetic
diffusion up to densities of $10^{10}$ cm$^{-3}$ or so, as many
models of this process suggest. The evidence suggests that most star
formation is not like this, however.

During the entire cycle of cloud evolution, a large fraction of the
molecular mass and a large fraction of the time are spent without
significant star formation, which is confined primarily to the
dense, short-lived inner regions. These dense regions form star
clusters, and they do this by forming protostars with relatively low
velocities that begin their lives inside gas filaments and
hierarchical subunits and mix over time inside the cloud core. The
outer parts may never get time to mix before core disruption. Then
they remain hierarchical up to scales of hundreds of parsecs until
galactic shear tears them apart.

The IMF would seem to be different for the three modes of star
formation discussed here: turbulence compression promotes scale-free
hierarchical structure while supercritical collapse promotes fast
relative motions and large-scale accretion. Considering IMF
simulations currently available, the supercritical cluster cores
could form proportionally more massive stars, thereby contributing
to mass segregation and a slight flattening of the IMF in dense,
supermassive clusters.

\acknowledgements I am grateful to the referee, Mark Krumholz, for
useful suggestions, and to Jonathan Tan and Yancy Shirley for
comments on the manuscript.

\clearpage


\end{document}